\documentclass{llncs}
\usepackage[utf8]{inputenc}
\usepackage[acronym]{glossaries}
\usepackage{graphicx}
\usepackage{booktabs}
\usepackage{color}

\title{Engineering Self-adaptive  \\Authorisation Infrastructures}
\author{Lionel  Montrieux\inst{1}
\and Rog\'{e}rio de Lemos\inst{2,3}
\and Chris Bailey\inst{2}
}

\institute{National Institute of Informatics, Japan
\and University of Kent, UK
\and CISUC, University of Coimbra, Portugal
\and The Open Univeristy, UK
}

\newacronym{abac}{ABAC}{Attribute-Based Access Control}
\newacronym{cia}{CIA}{Confidentiality, Integrity, Availability}
\newacronym{dac}{DAC}{Discretionary Access Control}
\newacronym{mac}{MAC}{Mandatory Access Control}
\newacronym{rbac}{RBAC}{Role-Based Access Control}

\begin{document}
\sloppy
\maketitle


\begin{abstract} 
As organisations expand and interconnect, authorisation infrastructures become increasingly difficult to manage.
Several solutions have been proposed, including self-adaptive authorisation, where the access control policies are dynamically adapted at run-time to respond to misuse and malicious behaviour.
The ultimate goal of self-adaptive authorisation is to reduce human intervention, make authorisation infrastructures more responsive to malicious behaviour, 
and manage access control in a more cost effective way.
In this paper, we scope and define the emerging area of self-adaptive authorisation by describing some of its developments, trends and challenges.
For that, we start by identifying key concepts related to access control and authorisation infrastructures, and provide a brief introduction to self-adaptive software systems, which provides the foundation for investigating how self-adaptation can enable the enforcement of authorisation policies.
The outcome of this study is the identification of several technical challenges related to self-adaptive authorisation, which are classified according to the different stages of a feedback control loop.
\end{abstract}


\section{Introduction}
\label{sec:intro}

A critical concern for organisations surrounds the assurances of confidentiality, integrity, and availability of their computer based resources.
To provide such assurances, organisations utilise access control to protect against unauthorised access.
Regardless of adopting a fine grained approach to access control, abuse of access is still possible.
Any form of access, no matter how restrictive, presents the risk of attacks due to uncertainty in user behaviour.
To accommodate for this risk, organisations employ a range of methods~\cite{Moore:2012} to monitor and audit access within their systems and resources.

Traditionally, human administrators are relied upon to actively identify and drive changes in access control in response to detected abuse, natural organisational change, or identified errors in the criteria for access.
It is challenging for human administrators to maintain a true awareness of the configuration of access, particularly within a run-time environment.
With no complete view of access, obtaining assurances~\cite{Hu:2011} against changes made to mitigate the abuse of access is limited.
This potentially enables erroneous changes that cause a greater impact to the organisation over identified abuse.
In addition, and as evident in case studies of historic insider attacks~\cite{Cappelli:2012}, the use of human administrators alone is inefficient in mitigating abuse in a timely manner.
Improving on access control methodologies is one solution, yet such approaches~\cite{Benantar:2005,janicke2012,mcgraw:2009,Park:2004} are unable to actively mitigate abuse, since they are constrained to a static definition of the criteria for access control at run-time.

Implementations of authorisation infrastructures~\cite{Chadwick:2008} must be capable in handling the dynamic aspect of risk at run-time, driven by the uncertainty in user behaviour.
It is therefore necessary for such systems to actively observe how access rights are being used, in order to infer whether the current criteria and assignment of access are enabling a user to conduct malicious activity.
A promising solution for the provision of dynamic support to authorisation infrastructures is the incorporation of self-adaptation. 

Self-adaptive systems are systems that are able to modify their behaviour and/or structure in response changes that occur to the system itself, its environment, or even its goals~\cite{Lemos:2013}.
Applying self-adaptive techniques to authorisation infrastructures enables the infrastructure to observe, reason, and act on its own configuration of access control.
Through the use of a feedback loop~\cite{Brun:2009}, it is possible to employ a clear separation of concerns between the decision for access, and decision for a management change, therefore reducing the complexity in the criteria for access that dynamic access control approaches introduce. 

The main contribution of this paper is identification of several technical challenges associated with the self-adaptation of authorisation infrastructures.
The relevance of the identified challenges is discussed in the context of insider threat examples related to a Customer Energy Management System (EMS) case study.
Another contribution is related to how the self-adaptation of authorisation infrastructures should be structured in order to handle parametric adaptations, i.e., the specification of access rights of subjects to resources, and structural adaptations, i.e.,  the enforcement of those specifications by controlling the subject's access to a resource.

The rest of paper is organised as follows.
Section~\ref{sec:background} presents the concepts and terminology related to access control and authorisation infrastructures, and provides a brief introduction to self-adaptive software systems.
In Section~\ref{sec:case-study}, we introduce a simple case study, based on NIST Smart Grid specification, that will be used as a basis for introducing self-adaptive authorisation infrastructures.
We review the related work on dynamic access control in context of self-protection in Section~\ref{sec:relatedWork}.
In Section~\ref{sec:dynamicAC},  we map our perception of self-adaptive authorisation infrastructures into the modelling dimensions for self-adaptive software systems.
Section~\ref{sec:techniques} identifies, in terms of the key stages of a feedback control loop, like the MAPE-K loop, some challenges associated with the engineering self-adaptive authorisation infrastructures.
Finally, Section~\ref{sec:conclusions} concludes the paper and indicates directions for future work.

\section{Background}
\label{sec:background}

The focus of this paper is the application of self-adaptation in the management of privileges and the rendering of access control decisions, in order  to reduce the need for human intervention, whilst reducing cost, and enabling systems to robustly adapt when responding to change.
As such, the following section discusses some background topics, including, prominent access control models, authorisation infrastructures as a means to implement access control,  static and dynamic access control, self-adaptive authorisation infrastructures that are capable of adapting themselves at run-time, and finally, insider threats that we employ as a motivation for managing access control.

\subsection{Access Control Models}
\label{sec:auth}

In the literature, the terms authorisation and access control are sometimes used interchangeably.
In this paper, we define them as follows.

\begin{definition}[Authorisation]
Authorisation refers to the specification of whether a subject has access to a resource.
\end{definition}

\begin{definition}[Access control]
Access control refers to the enforcement of authorisation by controlling (i.e., granting or denying) subject's access to a resource.
\end{definition}

The goal of access control is to prevent unauthorised access to protected resources.
A resource could be anything from a software system (e.g., web application and database) to an electronic device (e.g., electronic door lock and mobile phone).
Through the specification of authorisation, captured in terms of policies, an organisation garners a certain level of protection from unwanted access.


Authorisation embodies two concepts: identities and permissions.
An \textit{identity} is a digital representation of a subject (a user), where a subject could be a human being, a system, or even a process~\cite{Benantar:2005}.
An identity contains information about the subject, particularly relevant for authentication~\cite{Pashalidis:2003}, where a subject must identify themselves, for example, entering in a username and password, or use of biometrics~\cite{Ratha:2000}.
Most importantly, an identity contains a set of the subject's access rights (also referred to as privileges~\cite{Chadwick:2002}).
Access rights, as the name suggests, represent a subject's right of access to a resource, used in accordance to a set of \emph{permissions}.
Once a subject obtains the required access right(s) to a resource, the subject is said to be authorised.

\textit{Access control models} classify and define how permissions are expressed, who can define permissions, and what an access right looks like~\cite{saml}.
For example, Mandatory Access Control (MAC)~\cite{TCSEC} enables subjects with a set of security attributes to access resources in conformance to centrally specified permissions (i.e., defined by security administrators).
In contrast, Discretionary Access Control (DAC)~\cite{TCSEC} enables subjects in a similar sense to MAC to access resources, however, permissions can be specified by the subjects themselves in relation to the resources that they own.
Another access control model is the Bell-LaPadula Model (BLP)~\cite{Bell:2005} where permissions are based on labelled classifications, such as, \emph{Top Secret} or \emph{Public}, and a subject's level of security clearance.

Arguably, the most adopted access control model in industry is the Role-Based Access Control (RBAC) model~\cite{RBAC04}, where recently 50\% of the 150 companies surveyed by the National Institute of Standards and Technology (NIST) had adopted RBAC by 2010~\cite{OConnor:2010}.
RBAC introduced the notion of roles, whereby a role is assigned a set of permissions that enable access to a resource.
Finally, the Attribute-Based Access Control (ABAC) model~\cite{Yuan:2005} presents a more generic view of the RBAC model, where instead of roles, attributes (a type - value tuple) are used in order to collate and assign permissions.

\subsubsection{Role Based Access Control (RBAC).}
\label{sec:rbac}

The Role Based Access Control (RBAC) model is the culmination of work by Ferraiolo et al.~\cite{Ferraiolo:1995} and Sandhu et al.~\cite{Sandhu:1996} that led to the NIST RBAC standard~\cite{RBAC04}.
The RBAC standard is defined by three layers, each layer extending the layer prior with additional features.
These layers are referred to as Core, Hierarchical, and Constrained.

RBAC \emph{Core} defines the fundamental elements that must exist within an implementation of RBAC model, namely: subjects (identities), roles, resources, actions, permissions, and sessions.
\emph{Subjects} are assigned a set of roles, where a \emph{role} defines a function within an organisation (e.g., operations manager).
\emph{Roles} are assigned permissions, where each \emph{permission} details the ability for a \emph{subject} to execute an \emph{action} (e.g., print) on a \emph{resource} (e.g., printer).
A subject's \emph{session} captures a set of roles that the subject has currently activated.
RBAC \emph{Hierarchical} extends RBAC Core by introducing the ability for roles to inherit permissions of another role.
RBAC \emph{Constrained} extends RBAC Core and RBAC Hierarchical by introducing constraints in regards to subjects and roles.

A limitation of the RBAC model is the focus on roles as access rights, which restricts the ability of a subject to access resources only via the subject's organisational role(s).
This both limits or overly exposes access to a resource since roles lack the granularity to final control access to resources.
Potentially, organisations may have to create fictitious roles, not representative of the actual organisational structure, to ensure proper access to resources.
In addition, the RBAC model provides no means to address multi-organisational access control, where access control is managed between several organisations.
As a result, this could increase the complexity of roles and permission assignments within RBAC rules, making access more challenging to support.

Many proposals extend RBAC, highlighting not only RBAC's popularity in industry, but also in research.
Kalam et al. extend RBAC to include the notion of organisations in Organisation Based Access Control (OrBAC)~\cite{Kalam03}.
Introducing \emph{organisations} enables the specification of RBAC rules relevant to an organisation, where there are many sources of authority (SOAs) sharing access, and enabling organisations (and SOA) to define permissions solely for their own resources.
Similar work by Demchenko et al. also address the problems caused by multiple sources of authority, proposing Role Based Access Control for Distributed Multidomain Applications (RBAC-DM)~\cite{Demchenko:2007}.
Demchenko et al. highlights limitations of RBAC in collaborative environments (containing multiple SOAs), and addresses them via the use of multi-domain authorisation sessions (where an RBAC \emph{session} can span across several organisational domains).
Lastly, Bertino et al.'s GEO-RBAC~\cite{Damiani07} introduces the notion of \emph{location}, where a subject's geographical location influences the activation of a subject's assigned roles.
GEO-RBAC addresses the need for spacial aware access control, where subjects may only access a resource depending on their location.

\subsubsection{Attribute-Based Access Control (ABAC). }
\label{sec:abac}

Attribute Based Access Control (ABAC) is a recent development in access control models.
There are a number of proposals~\cite{Hu:2013}, critiques~\cite{Sandhu12a}, and implementations~\cite{Chadwick:2008,shibboleth:2004,xacml}.
ABAC can be considered a natural progression from the RBAC model, whereby instead of permissions assigned to roles, permissions are assigned to attributes of a subject, resource, and their environment.
An attribute describes some aspect of their associated entity, such as, a name or user group (for a subject), the number of active sessions (of a resource), or time of day and location (in the system's environment).
These attributes can be defined in such a way to create permissions with a fine granularity of access, where a permission may state that subjects only from user group `HR' can access a resource with no more than 10 active sessions, between the hours of 9am and 5pm.
In addition, ABAC is seen as a generalisation of RBAC, where RBAC roles are implemented as ABAC attributes assigned to subjects.


ABAC implementations and proposals have put forward additional criteria for access control, as opposed to simply replacing the notion of roles in RBAC with attributes.
Notably, environment conditions are considered in order to provide additional context to a subject's request for access.
For example, a subject requesting access outside of normal office hours should not be granted access, despite having the necessary attributes to gain authorisation to the resource.
The inclusion of environment conditions has the ability to expand the criteria necessary to award access, and further protect an organisation's resources from attacks (e.g., credential stealing attacks~\cite{Sharma:2010}, by blacklisting IP addresses based on location data observed in the environment).

Sandu has argued that the leap from RBAC roles to the use of attributes offers a number of benefits~\cite{Sandhu12a}, highlighting the fact that ABAC unifies many access control models, for example, roles (RBAC), location (GEO-RBAC), security labels (Bell-LaPadula), and access control lists (DAC).
However, the resulting benefits of ABAC come with increased complexity.
Organisations now have to be more specific when utilising ABAC, as access rights could be represented as anything that might be owned by a subject, resource, or environment.
This has the potential to lead to conflicts, or increased challenges when managing access, due to no clear representation of an access right.

\subsubsection{Implementing Access Control Models.}
\label{sec:xacml}

Traditionally, access control models have been implemented as bespoke components of information systems.
Implementation concerns both `authorisation' being how to capture and express identities, privileges, and permissions, and `access control', referring to the process that can render access control decisions.
A problem with this approach is the heterogeneous qualities of resources an organisation may wish to protect, often requiring each resource (e.g., a web application) to implement its own form of access control.

A solution to this problem is implementing access control models in a service orientated way, as demonstrated by the eXtensible Access Control Markup Language~\cite{xacml}.
XACML is a popular standard for implementing ABAC and RBAC models, and provisions a reference architecture in which to guide implementation.
XACML standardises the way in which identities and permissions are defined, communicated, and assessed in order for its reference architecture to render access control decisions.
It does this through the use of authorisation policies (to express identities, privileges, and permissions as `attributes' and `rules'), and the use of standardised protocols (e.g., SAML~\cite{saml}).
Authorisation policies embody the `authorisation' aspect of an access control model, whereas the protocols support `access control' via retrieval of privileges, and deliverance of access control decisions.

XACML's reference architecture describes a set of components that exist to facilitate access to protected resources.
The reference architecture defines a four tier process to access control: 
\emph{Enforce} requests and decisions to access,
\emph{Decide} upon access,
\emph{Support} retrieval of credentials and policies,
and \emph{Manage} administration of policies.
This process is implemented through a set of conceptual components that when combined achieves access control (Table~\ref{tbl:xacmlComponents}).
These components are the enabling factors for controlling access, whereby in real systems that implement such components, access control can easily be monitored and managed.
A key selling point of the XACML reference architecture is the separation between access control and resources, where access control primarily becomes a service that resources and users can rely upon.

\begin{centering}
\begin{table}
\begin{tabular}{lllllll}
\toprule
Component & Description \tabularnewline
\midrule
Policy enforcement point (PEP) & Makes access requests and enforces \\ & access decisions \tabularnewline
Policy decision point (PDP) & Evaluates access requests against policies \\ & to provide access decisions \tabularnewline
Policy information point (PIP) & Contains subject identity information \\ & (attributes) \tabularnewline
Policy retrieval point (PRP) & Contains ABAC authorisation policies  \\ & to govern access decisions \tabularnewline
Policy administration point (PAP) & The source of authority / system that \\ & issues access control policies \tabularnewline

\end{tabular}
\caption{XACML components}
\label{tbl:xacmlComponents}
\end{table}
\end{centering}

The XACML reference architecture has arguably sparked the rise of access control as a service, where we refer to such solutions as \emph{authorisation infrastructures}.

\subsection{Authorisation Infrastructures}
\label{sec:authinfrastructures}


An authorisation infrastructure~\cite{Chadwick:2008} is a loose term for a collection of services and mechanisms that implement an access control model.
There are a number of varying terms for authorisation infrastructures, such as, the ones defined by authentication and authorisation infrastructures (AAIs)~\cite{Lopez:2004}, XACML's reference authorisation architecture~\cite{xacml}, and privilege management infrastructures~\cite{Chadwick:2002}.
We adopt the following rather simple definition for authorisation infrastructure.


\begin{definition}[Authorisation Infrastructure]
Authorisation infrastructures facilitate the management of identities, privileges and policies, and render access control decisions.
\end{definition}



The key facet of authorisation infrastructures is the use of services that provide access control external to an organisation's resources.
This implies a separation of duties between provisioning of services by the resources, and the assessment of right to access~\cite{Chadwick:2008, Lopez:2004, xacml}.
Specifically, access control is implemented by the following key services:
\begin{description}
\item [\textit{Identity services}] responsible for the the management of subject access rights, such as, access rights and subject identifiers.
\item [\textit{Authorisation services}] responsible for the evaluation of access rights against access control rules, and decision of access.
\end{description}


The combination of both identity services and authorisation services should conform to an  access control model (e.g., RBAC~\cite{RBAC04}).
Based on existing implementations~\cite{Chadwick:2008, Lopez:2004, xacml}, identity services may authenticate subjects, and maintain, assign and release a subject's access rights (i.e., privileges) to authorisation services based on policies (e.g., Shibboleth's attribute release policy~\cite{shibboleth:2004}).
Examples of an identity service include directory services, such as, the Lightweight Directory Access Protocol (LDAP)~\cite{Koutsonikola:2004}.
Other forms of identity services include credential issuing services (such as, SimpleSAMLphp~\cite{simpleSAML} and the Shibboleth identity provider~\cite{shibboleth:2004}).
These types of identity services not only maintain a subject's access rights (privileges), but can be configured to decide what access rights can be issued and released to given services across multiple domains.
Authorisation services may validate and evaluate a subject's access rights against a set of policies (e.g., PERMIS's access control and credential validation policies~\cite{permis}).
Examples of authorisation services include, the axiomatics policy server~\cite{axiomatics}, PERMIS standalone authorisation service~\cite{permis} (both of which utilise the XACML standard to define access control policies), and the community authorisation service (CAS)~\cite{Pearlman:2002}.

\begin{figure}[!htbp]
\centering
\includegraphics[width=0.9\textwidth]{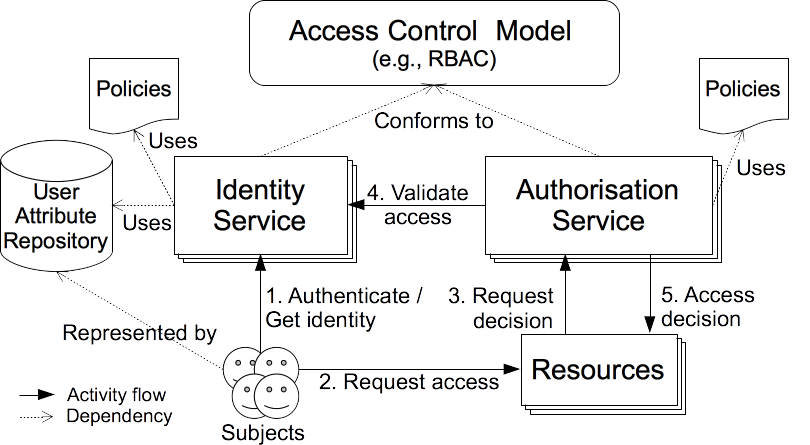}
\caption{General authorisation infrastructure model}
\label{fig:appDomainConcept}
\end{figure}

Figure~\ref{fig:appDomainConcept} defines a general model of an authorisation infrastructure that abstracts away from its varying implementations.
With reference to the flow of communication to obtain authorisation, subjects (users) authenticate  (1) with a given identity service that maintains a set of access rights for each subject.
The authenticated subject can then request (2) access to a particular resource.
The resource's policy enforcement point (PEP) communicates with an authorisation service (3), which can first validate (4) the subject's set of access rights, and then decide upon access.
The authorisation service sends a response back to the PEP with a message indicating whether authorisation should be granted or denied (5).


As already mentioned, authorisation infrastructures rely on policies to derive access control decisions.
Authorisation infrastructures may utilise a variety of policy types, where an instance of a policy type will express rules relevant to a particular service within an authorisation infrastructure.
For example, policies within authorisation services are used to define the constraints of access (i.e., RBAC role permission assignments), whereas policies and subject attribute repositories (e.g., LDAP~\cite{Koutsonikola:2004}) within identity services contain or define what subjects have in terms of assigned access.
With this in mind, there are four types of authorisation policies, which are defined as follows.
\begin{description}
\item [\textbf{Access Control Policy}.]An access control policy specifies the security controls of what credentials a subject must own in order to gain access to a set of protected resources, what obligations they must conform to, and what conditions they must meet.
\item [\textbf{Credential Validation Policy}.] A credential validation policy defines what credentials an identity service is trusted to issue.
\item [\textbf{Delegation Issuing Policy}.] A delegation issuing policy defines the trust in the extent of access a subject can delegate unto others.
\item [\textbf{Credential/Attribute Release Policy}.] A credential/attribute release policy defines what information an identity service will release on behalf of a subject to any requesting authorisation services or resources.
\end{description}

Associated with policies and access rights is the notion of source of authority (SOA) and issuer~\cite{Chadwick:2008}.
A \emph{source of authority} is the owner of a resource that establishes the rules of access (as policies) to their resources.
An \emph{issuer} is the identity service or person responsible for issuing to a subject a set of access rights, which are either stored in an attribute repository as unsigned or signed attributes~\cite{x509}, or are generated at time of request~\cite{saml}.

Lastly, an additional quality of authorisation infrastructures is the ability to operate in federated environments (i.e., components of an authorisation infrastructure become component systems managed across multiple organisations).
This is often referred to as federated identity management, or federated access control~\cite{Demchenko:2007,Hu:2013,Kalam03,Thompson:1999}, and enables the sharing of access across multiple management domains (organisations).
Various access control models are suitable for federated access control, demonstrated by several implementations~\cite{Chadwick:2008,shibboleth:2004,simpleSAML}.

\begin{figure}[!h]
\centering
\includegraphics[width=.85\textwidth]{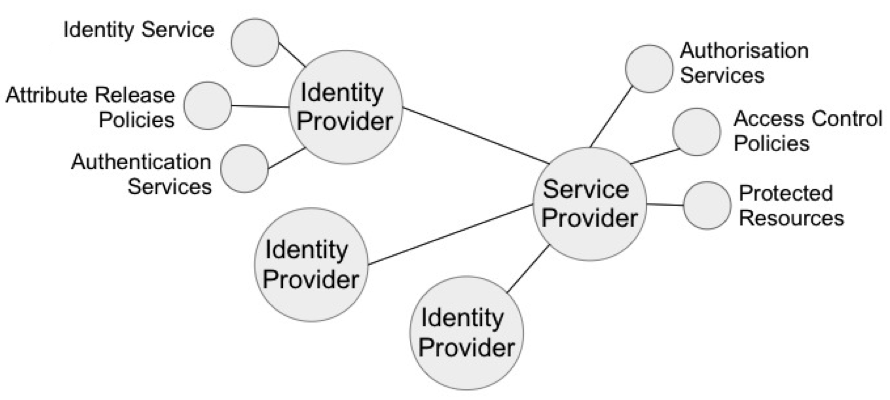}
\caption{Conceptual view of a federated authorisation infrastructure}
\label{fig:federated_access}
\end{figure}

Figure~\ref{fig:federated_access} conveys a high level overview of a federated environment, containing a service provider (SP) organisation and several identity provider (IdPs) organisations.
An SP organisation offers access to their protected resources, whereas IdP organisations consume access to those protected resources.
Subject identities managed by an identity provider component system can be assigned a set of attributes that are stored within an identity service (e.g., simpleSAMLphp~\cite{simpleSAML}).
Subjects can use their attributes to gain access to a SP's resources given that the SP trusts the IdP.
To control the release of attributes, some IdPs may define attribute release policies~\cite{shibboleth:2004} to prevent certain types of information from being released to service providers.
The service provider ultimately decides upon access through the use of authorisation services (which provide access control decisions local to the organisation).

\label{ssec:federatedAccessControl}

\subsection{Static and Dynamic Access Control}
\label{sec:staticDynamicAC}

We have seen how access control is a key element when implementing authorisation infrastructures.
However, one thing not addressed is the distinction between traditional (static) approaches to access control, to more recent (dynamic)  advanced approaches.
In a static approach to access control, a user's access rights are assessed against a set of security controls in order to determine access (e.g., RBAC~\cite{RBAC04}).
This is limited since at time of access no additional context is assessed, such as, the user's historical access, their location, time of day, or other external factors.
With this in mind, static approaches are presumptuous in that, should a user have the necessary access rights, they should be awarded access.

Arguably, it is not always the case that access should be granted despite the user owning the necessary access rights.
As such, there is a growing focus on dynamic approaches to access control that allow organisations to define a finer grain of control over access in response to varying risks, threats, and environment states.
The definitions for static and dynamic access control are as follows.

\begin{definition}[Static Access Control]
Static access control refers to the evaluation of a subject's access rights against a set of immutable authorisation policies for deciding the subject's access to a resource, regaess of the context in which the request is made and evaluated.
\end{definition}

\begin{definition}[Dynamic Access Control]
Dynamic access control  refers to the evaluation of a subject's access rights against a set of authorisation policies for deciding the subject's access to a resource, taking into account the context in which the request is made and evaluated.
\end{definition}


Dynamic access control differs from static access control because it is capable of employing various security controls that are related to changes in the state of the environment or protected resources, and user activity.
As such, an authorisation policy may contain a diverse set of access control rules to accommodate a wide variety of scenarios (e.g., a rise in national security threat levels~\cite{mcgraw:2009}).
Appropriate access control rules are applied to requests for access in a mutually exclusive manner, given the context (i.e., state of the environment, such as, user activity or time of day) that surrounds the request.

The overall goal of dynamic access control is to reduce human intervention, make access control more responsive to attacks, and more cost effective.
Several techniques have been proposed, including, resource usage~\cite{Park:2004}, temporal properties~\cite{janicke:2012}, risk~\cite{mcgraw:2009}, and trust~\cite{Bistarelli:2010,Serrano:2009}.
For example, in usage control~\cite{Park:2004} a perception of user activity is maintained over time and evaluated against thresholds of usage (e.g., a staff member may not print more than 100 pages per day), alongside traditional access control rules (e.g., a user must be assigned the role of staff to print).
Additionally, ABAC can be seen as a dynamic access control model given its ability to define permissions that can be valid for a multitude of system states.


\subsection{Self-adaptive Authorisation Infrastructures}
\label{sec:selfadaptiveAI}


With the goal of reducing human intervention, self-adaptation can be incorporated into existing authorisation infrastructures, thus enabling these infrastructures to manage themselves, at run-time, the definition of authorisation policies and process of access control.
In particular, the focus of this paper is how self-adaptation can be integrated with authorisation infrastructures, and how authorisation infrastructures can self-protect against insider threats.

\subsubsection{Self-adaptation.}
\label{selfadaptation}

Self-adaptation enables a system to adjust itself in response to
changes that might affect itself or its environment.
Self-adaptive systems can be defined as follows.

\begin{definition}[Self-Adaptive Systems~\cite{Lemos:2013}]
``Systems that are able to modify their behaviour and/or structure in response changes that occur to the system itself, its environment, or even its goals.''
\end{definition}

\begin{figure}[!hbtp]
\centering
\includegraphics[width=.55\textwidth]{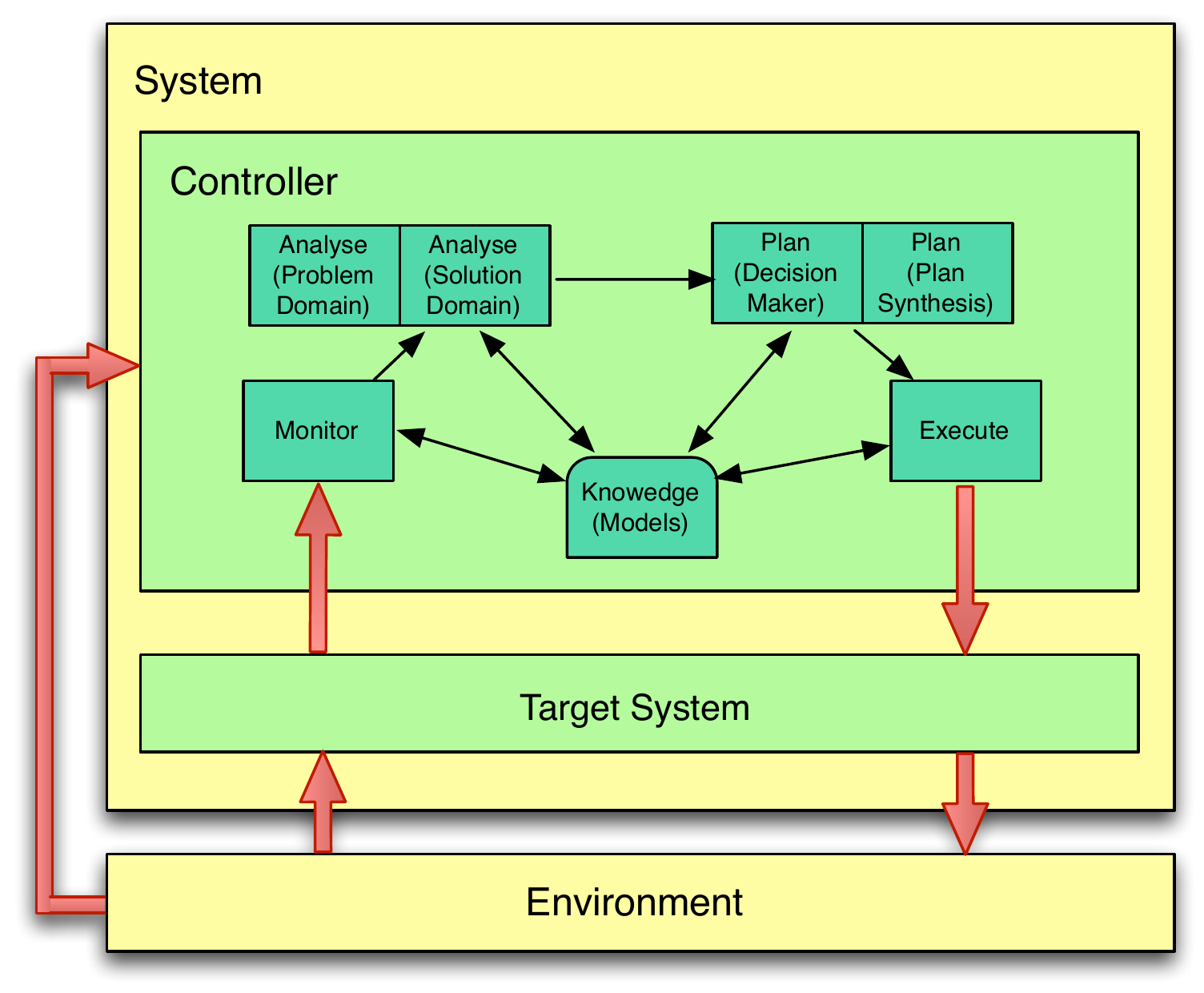}
\caption{Modified version of the MAPE-K reference model for autonomic computing~\cite{Kephart:2003}.}
\label{fig:mapek}
\end{figure}

Although there are several reference models for self-adaptive systems~\cite{Kephart:2003,Kramer:2007,Oreizy:1999}, most of them share the common use of a feedback loop~\cite{Brun:2009,Dobson:2006,Hellerstein:2004}.
In this paper, we adopt as a feedback control loop, the Monitor, Analyse, Plan, Execute - Knowledge (MAPE-K) reference model~\cite{Kephart:2003}, as shown in Figure~\ref{fig:mapek}.
In this diagram, the main feedback control loop, which embodies the stages of the MAPE-K reference model, observes (via probes) and adapts (via effectors) a target system.
The \emph{Monitor} stage enables to obtain the state of the target system and its environment.
The \emph{Analyse} stage analyses the state of the target system and its environment in order, first, to decide whether adaptation should be triggered (\textit{Solution Domain}), and second, to identify the appropriate courses of action in case adaptation is required (\textit{Problem Domain}).
The \emph{Plan} stage, first, selects amongst alternative course of action those that are the most appropriate (\textit{Decision Maker}), and second, generates the plans that will realise the selected course of action (\textit{Plan Synthesis}).
The \emph{Execute} stage executes the plans that deploy the course of action for adapting the system.
Finally, \emph{Knowledge} represents any information related to the perceived state of the target system and environment that enables the provision of self-adaptation.

Applying the MAPE-K reference model, we view an authorisation
infrastructure as the target system, and all the rest, including the users and protected resources, as the environment.
The role of a controller\footnote{Also referred to as the self-adaptive layer.} seeks to monitor both the target system and the environment in which to drive changes at run-time within the authorisation infrastructure.
With this in mind, self-adaptation is capable of extending traditional approaches to access control, where such approaches become capable to respond to unplanned states, evolve to changing user needs, and maintain assurances in confidentiality, integrity, and availability of resources.

\subsubsection{Self-adaptive authorisation and self-adaptive access control.}
\label{selfauthaccess}

Self-adaptive authorisation has already been proposed by Bailey et al~\cite{Bailey:2011,Bailey:2014,baileyPhD2015}, where legacy based authorisation infrastructures have been shown to mitigate,  at run-time, attacks via the adaptation of authorisation (e.g., adaptation of authorisation policies and subject privileges).
We define self-adaptive authorisation as follows.

\begin{definition}[Self-Adaptive Authorisation]
Self-adaptive authorisation refers to
the run-time adaptation of the specification of whether a subject has
access to resources.
\end{definition}

The incorporation of self-adaptation into authorisation has highlighted a number of challenges that this paper aims to address, including the engineering of self-adaptive authorisation infrastructures, and  practicalities of operating such systems at run-time.
First, it is important to identify the differences between static approaches to access control (i.e., traditional, such as, RBAC), dynamic approaches (i.e., adaptive, such as, risk based), and self-adaptive ones.

Let us consider a subject requesting access to a resource outside of normal working hours, who then abuses such access in order to jeopardise the confidentiality of a resource.
A static approach (i.e., static access control) will evaluate access based on purely the subject's access rights alone, without considering the time of day, or the subject's activity.
A dynamic approach (i.e., dynamic access control) may select, from a pre-existing set of access control rules, a rule applicable for that time of day, using environmental attributes and the subject's access rights.
On the other hand, a self-adaptive approach may, at run-time, generate, modify, or remove the active set of access control rules (e.g., deploying a new authorisation policy, or revoking a set of user access rights) should a user be detected while abusing their access rights outside of normal working hours.
Additionally, modifications instructed by a self-adaptive approach are based on a maintained perception of state of its target system and its environment.

Self-adaptive authorisation alone has some limitations.
Specifically, it is limited to only mitigating attacks (e.g., insider threats) within the boundaries of an authorisation infrastructure's implemented access control model, where adaptation is primarily parametric.
Should services of an authorisation infrastructure suffer an attack, or the implemented access control model becomes vulnerable, an additional scope of adaptation is needed.
As such, it is important to address the possibility of self-adaptive access control, which we define as follows.


\begin{definition}[Self-Adaptive Access Control]
Self-adaptive access control refers to the run-time adaptation of the enforcement of authorisation by controlling the subject's access to a resource.
\end{definition}


\begin{figure}[!htbp]
\centering
\includegraphics[width=0.7\textwidth]{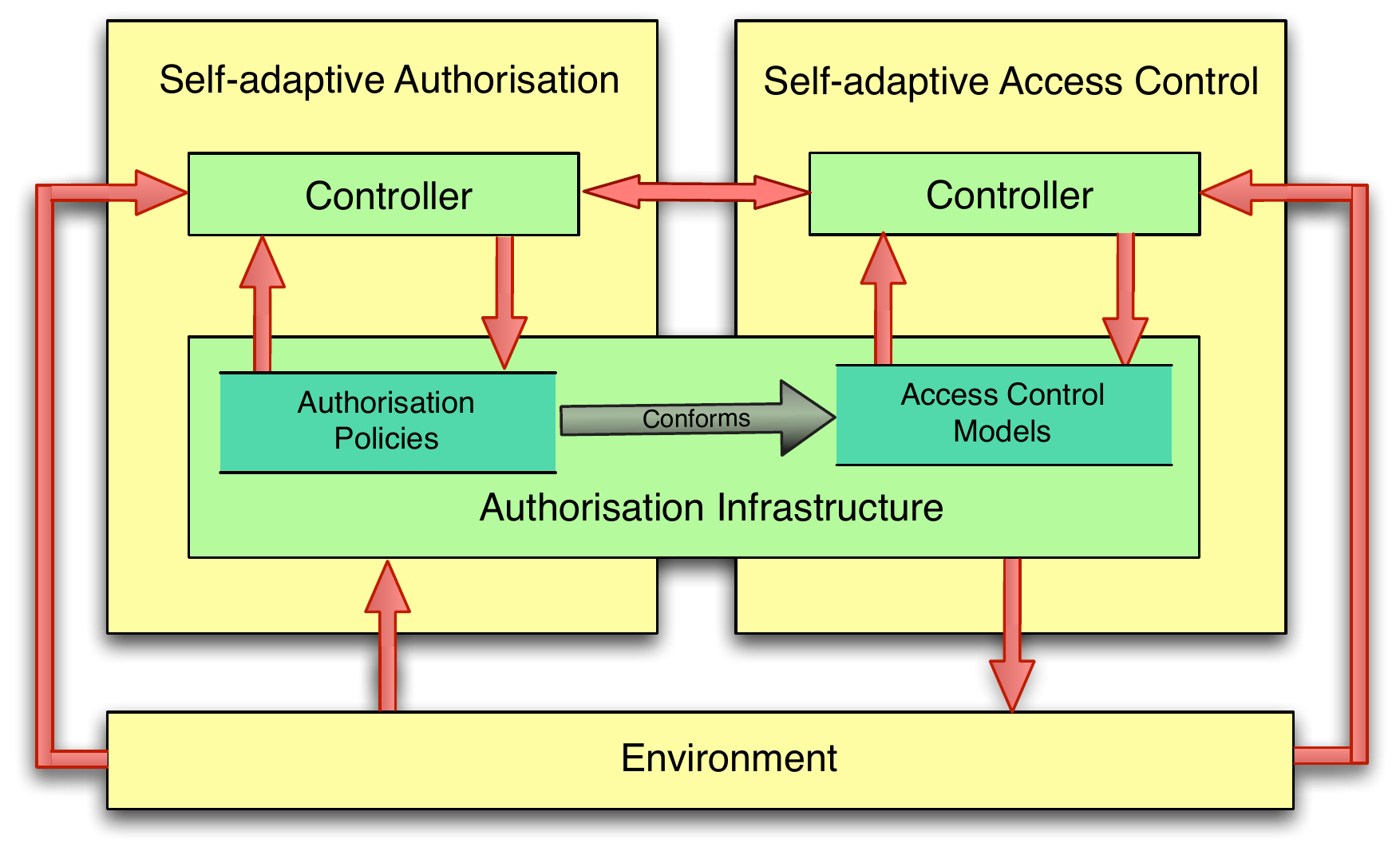}
\caption{Self-adaptive authorisation and self-adaptive access control}
\label{fig:saaz-saac}
\end{figure}

Figure~\ref{fig:saaz-saac} emphasises the marriage of self-adaptive authorisation and self-adaptive access control, which allow us to mitigate attacks more effectively and efficiently, depending on the type of attack observed.
Authorisation being the collection of policies that govern access, and access control being the process in how an access decision is achieved.
From the diagram, we can see a distributed control topology of two controllers operating together in mitigating potential attacks originating from the  environment of the authorisation infrastructure.
The controller associated with self-adaptive authorisation observes activity of the authorisation infrastructure and its environment in order to gain a perception of malicious behaviour with relevance to the current state of authorisation policies.
Should malicious behaviour be observed, this controller can adjust deployed authorisation policies to mitigate attacks.
Similarly, the controller associated with self-adaptive access control may observe the authorisation infrastructure and its environment in order to identify if the current state of the employed access control model is fit for purpose.
For example, external threats may warrant additional steps in validating subject credentials, and as such, the controller may deploy credential validation services~\cite{Chadwick:2008} between policy decision and policy enforcement points~\cite{xacml}.
Based on the above, we define self-adaptive authorisation infrastructures as follows.

\begin{definition}[Self-Adaptive Authorisation Infrastructure]
Self-adaptive authorisation infrastructures refer to the run-time adaptation of the collection of authorisation policies and their enforcement.
\end{definition}


\subsubsection{Self-protection.}
\label{self-protection}

Self-protection is of particular relevance since the goal of  this work is to manage access control in order to mitigate abuse of access.
Self-protecting systems can be defined as follows.

\begin{definition}[Self-Protecting System~\cite{yuan:2014}]
``Self-protecting systems are a class of autonomic systems capable of detecting and mitigating security threats at run-time.''
\end{definition}

There are various self-protective solutions that seek to detect and mitigate malicious behaviour.
However, few works exist that are able to concretely address self-protection with a view to mitigate the abuse of access.
Whilst many systems appear to be self-protective, such as, intrusion response systems~\cite{mu:2010,stakhanova:2007, Strasburg:2009}, many are only adaptive and lack an awareness of `self'.
A self-protecting system is clearly demonstrated by Yuan et al.'s architectural based self-protection framework~\cite{yuan:2013}, where a system maintains a modelled state of its own system architecture in which to guide mitigation of threats.

\subsection{Insider Threats }
\label{sec:internal}

Insider threat refers to an organisation's risk of attack by their own users or employees.
It is fast becoming a prominent topic that organisations need to address, as highlighted by recent scandals in the media~\cite{BBC:2014,Booth:2010,Walsh:2006}.
This is particularly relevant to access control, where the active management of authorisation has the potential to mitigate and prevent users from abusing their own access rights to carry out attacks.
The CERT Guide to Insider Threats (Cappelli et al.)~\cite{Cappelli:2012} defines malicious insider threats as the following.

\begin{definition}[Insider Threat~\cite{Cappelli:2012}]
``A malicious insider threat is a current or former employee, contractor, or business partner who has or had authorised access to an organisation's network, system, or data and intentionally exceeded or misused that access in a manner that negatively affected the confidentiality, integrity, or availability of the organisation's information or information systems.''
\end{definition}

Capelli et al.~\cite{Cappelli:2012} classify three types of insider threat:
\emph{sabotage}, where malicious users attempt to damage or corrupt organisational resources,
\emph{theft} of intellectual property, where organisational resources are stolen and distributed,
and \emph{fraud}, where activity is covered up or information is used to commit crimes, such as, falsifying money transfers.

A common characteristic of insider threat is that malicious insiders utilise their knowledge of their organisation's systems, and their assigned access rights, to conduct attacks.
This places a malicious insider in a fortuitous position, whereby the insider (as an authorised user) can cause far greater damage than an external attacker, simply due to their access rights ~\cite{Caputo:2009}.
Such form of attack is representative of the attacks that many organisations consider to be most vulnerable from, being the abuse of privileged access rights by the employees of an organisation~\cite{vormetric:2013}.


Unless additional measures are put into place, malicious insiders can abuse existing security measures, where current approaches fail to robustly adapt and respond to the unpredictable nature of users.
For example, traditional approaches to access control assume that if a user has authenticated, and has the required access rights, access to resources should be given.
Whilst there are a number of novel techniques that enable the detection of insider threat~\cite{ibmThreatTool,nblgcww:2014,Spitzner:2003}, there is little research that utilises such techniques within an automated setting.
Many existing approaches require analysis by human agents to identify and execute resultant actions to mitigating attacks.

\section{Case Study}
\label{sec:case-study}

In this section, we present a case study to illustrate the challenges
in self-adaptive authorisation infrastructures, discussed in this paper.
The case
study is based on a subset of the NIST Smart Grid Cybersecurity specification~\cite{nist}.



\subsection{The NIST Smart Grid Cybersecurity Specification}
\label{sec:case:nist}

The NIST IR 7628 \emph{Guidelines for Smart Grid
  Cybersecurity}~\cite{nist} is an advisory report ``\emph{intended to
  facilitate each organization’s efforts to develop a cyber security
  strategy effectively focused on prevention, detection, response, and
  recovery}''~\cite{nistIntro}, in the context of the transformation
  of the US electricity system into a smart grid.
The report was
  released in three volumes: ``\emph{Smart Grid Cyber Security
  Strategy, Architecture, and High-Level
  Requirements}''~\cite{nistSecurity}, ``\emph{Privacy and the Smart
  Grid}''~\cite{nistPrivacy}, and ``\emph{Supportive Analyses and
  References}''~\cite{nistAnalyses}.

The first volume presents a high-level overview of the proposed
framework, together with high-level security requirements.
The report
identifies 49 actors (including energy providers, customers,
regulators, etc.) involved in the Smart Grid, and defines 22 logical
interface categories, over 7 domains.
The domains are transmission,
bulk generation, operations, distribution, marketing, service
provider, and customer.
Due to the wide scope of the specification
and its very high-level nature, we focused on a small subset of the
framework, and constructed a more detailed architecture and
requirements when necessary.

\subsection{Smart Grid Cybersecurity: A Self-adaptive Authorization
  Infrastructure}
\label{sec:case:case}

The case study is centred around the \emph{Customer Energy Management
  System} (EMS), an actor in the customer domain described as
  ``\emph{an application service or device that communicates with
  devices in the home.
The application service or device may have
  interfaces to the meter to read usage data or to the operations
  domain to get pricing or other information to make automated or
  manual decisions to control energy consumption more efficiently.
The
  EMS may be a utility subscription service, a third-party offered
  service, a consumer-specific policy, a consumer-owned device, or a
  manual control by the utility or customer}''~\cite{nistSecurity}. In
  this case study, we assume a certain level of automation for the
  EMS, as an entirely manual control would not fit the purpose of our
  work.

The framework defines only the logical interfaces between the EMS and other
actors in the customer, operations and service provider domains.
Therefore, the scope of our case study will be restricted to
those domains and actors, and only to the extend relevant to the
operation of the EMS.
The choice of the EMS, as the main focus of our
case study, stems from the ability to involve a variety of possible
scenarios involving several third-parties.
Since the EMS has to deal
with data flowing to and from those third-parties, as well as, to keep
the data safe from unauthorised access, it is a good candidate for a
case study focused on  self-adaptive authorisation infrastructures.

\subsubsection{Actors}
\label{sec:case:case:actors}

The following actors from the Smart Grid Cyber Security
Specification~\cite{nistSecurity} interact with the EMS, and are
relevant to the case study:

\begin{description}
\item[Customer Appliances and Equipment:] ``\emph{A device or
instrument designed to perform a specific function, especially
an electrical device, such as a toaster, for household use. An
electric appliance or machinery that may have the ability to be
monitored, controlled, and/or displayed.}''
\item[Customer Distributed Energy Resources (DER): Generation and Storage:] ``\emph{Energy
generation resources, such as solar or wind, used to generate and
store energy (located on a customer site) [...]}''
\item[Meter:] ``\emph{Point of sale device used for the transfer of
product and measuring usage from one domain/system to
another.}''
\item[Customer Premise Display:] ``\emph{A device that enables
customers to view their usage and cost data within their home or
business.}''
\item[Customer Information System:] ``\emph{Enterprise-wide software
applications that allow companies to manage aspects of their
relationship with a customer.}''
\end{description}

The EMS maintains authorisation policies that determine which actors
can access which data, and under which conditions. Throughout the
paper, the case study will be used to illustrate various
self-adaptation techniques that can be applied to the adaptation
authorisation policies.

\subsubsection{Goals}
\label{sec:case:goals}

The system protects sensitive information, and allows selected actors
to access some of that information. The goals of the system, at a high
level, are the typical CIA properties:

\begin{description}
\item[Confidentiality:] the system must guarantee that the data it
  protects cannot be used by a malicious subject. The system must also
  guarantee that legitimate users cannot use their privileges to
  access information that they are not meant to be able to access.
\item[Integrity:] the system must make sure that the data and commands
  it protects cannot be compromised.
\item[Availability:] the system must ensure that the data and commands
  are protected against malicious subjects that would try to make the
  system inaccessible.
\end{description}

The high level goals of the system are refined in the scenarios below
(Sections~\ref{sec:case:scenario-1} to \ref{sec:case:scenario-2}).

\subsubsection{Initial State of the System}
\label{sec:case:initial}

The system, in its initial state, involves the following actors:

\begin{description}
\item [Customer energy management system (EMS):] a self-adaptive software
  system, running in the household, owned and controlled by the
  couple, that interfaces with all the actors below. The system
  exposes an API that allows authorised applications and web services
  to access data produced by the other actors, or to issue commands to
  said actors.
\item [Customer appliances and equipment:] an air conditioning unit, a
  water boiler, a thermostat, and an CCTV camera. They are all
  connected to a local network.
\item [Customer Distributed Energy Resources (DER):] solar panels, that can produce electricity. The
  electricity used can either be used in the household, fed to the
  grid, or wasted.
\item [Meter:] a smart meter, that can be queried remotely.
\item [Customer premise display:] an application running on the couple's
  smartphones.
\item [Customer information system:] the software system, running on the
  energy retailer's infrastructure, that allows the household to
  access, in real-time, data on their energy consumption, as well as
  information of whether, and when, they are allowed to feed energy to
  the grid.
\end{description}

The default authorisation policies, deployed on the EMS, can be
summarised as follows:

\begin{itemize}
\item the EMS can read the temperature and humidity in the household
  from the thermostat;
\item the EMS can turn the AC unit on and off, and switch it between
  the cooling and heating modes;
\item the EMS can turn the water boiler on and off, and read the
  temperature of the water in the tank from the boiler's sensor;
\item the EMS can find out the location of each member of the
  household, by querying their smartphones' location service;
\item the EMS can read the real-time energy production from the solar
  panels;
\item the EMS can query the meter to find out the current energy
  consumption;
\item the EMS can feed data to the customer premise display;
\item no third party apps or services are allowed to read or write any
  information from or to the EMS;
\item the EMS can get real-time and historical data from the customer
  information system;
\item the Customer Premise Displays (CPD) app running on the couple's
  smartphones have read and write access to a range of data from the
  EMS, and can issue commands to the connected devices in the
  household;
\item the customer information system can access the household's meter
  data at all times.
\end{itemize}

\subsection{Scenarios}
\label{sec:case:scenarios}

\subsubsection{First Scenario: A Compromised Service}
\label{sec:case:scenario-1}


The first scenario deals with compromised third party services, and
the resulting risk of misuse of credentials. In this scenario, the
couple subscribes to an online service that uses their location to
automatically regulate the heating system in their home, as well as
the amount of hot water available. 
The application requires read
access to the couple's locations (through their smartphones' location
service), read access to the thermostat and water boiler data, and
write access to the AC/heating unit, and water boiler. By using the
service, the couple's goal is to minimise their energy bill, while
still enjoying hot water and room temperature between 19-21$^\circ$C
when home. The service should analyse the data it collects, and infer
the best times to turn the AC/heating unit and the water boiler on and
off.

The specific authorisation goals for this scenarios are the following:
\begin{itemize}
\item the system must ensure that the data is only collected by the
  third-party service in a secure way, through an encrypted channel,
  and using state of the art authentication protocols.
\item the system must monitor access to the data in order to detect
  any misuse. In particular, it must ensure that patterns
  of read and write requests do not suddenly change, which may
  indicate a compromised service.
\end{itemize}

We can envisage a scenario in which, after a few months on continuous
use, the service gets compromised, and starts issuing commands to the
water boiler and the AC/heating unit that do not achieve the
goals. 
Furthermore, getting location, water temperature and house
temperature reading allows the attackers to infer details about the
couple's habits and movements.
The erratic behaviour of the service is
detected, the service's credentials are temporarily revoked, and the
users are notified of the incident.

\subsubsection{Second Scenario: Intrusive Energy Retailer}
\label{sec:case:scenario-2}

In the second scenario, the energy retailer is potentially threatening
the couple's privacy. The company has read access to the couple's
meter for billing their energy use. However, if the energy retailer
queries the meter too often, it may be able to infer patterns about
the couple's life. By ``too often'', we mean more than necessary for
the operation of the service, and sufficiently to be able to make
conjectures that would threaten the couple's privacy, such as the
study of their movements or habits. 
The frequency of making a query threatens the couple's privacy for different reasons.
It could be the
energy retailer's policy to snoop on its customers, but it could also
be that the energy retailer's infrastructure was compromised, either
by an internal agent (such as a disgruntled employee) or by an
external entity (such as a competitor or a spying agency).

The specific authorisation goals for this scenario are the same as for the first scenario. Instead of the third party, the queries are made by the energy retailer.

As the system is running, the self-adaptive authorisation infrastructure eventually detects that the energy retailer queries the meter too often, and chooses to reduce the number of times the energy retailer can query the meter every day, in order protect the couple's privacy. 
The users are also notified of the incident.

\subsubsection{Third Scenario: Data Deletion}
\label{sec:case:scenario-3}

The third scenario deals with integrity issues.
In this scenario, the authorisation infrastructure monitors the use of the Customer Premise Displays (CPD) app.
Part of the CPD app functionality is to allow the couple to selectively delete some data, such as their location at a certain point in time, either for privacy reasons, because of errors in the location data, or because they do not want some unusual data points to be taken into account in the decisions taken by the third party service described in the first scenario.

The specific authorisation goal for this scenario is the following extension of the integrity goal:

\begin{itemize}
\item the authorisation infrastructure must prevent malicious data destruction, by monitoring the usage of the CPD apps for pattern of suspicious behaviour, such as the deletion of vast quantities of data within a short time frame.
\end{itemize}
Once suspicious behaviour is  detected by the authorisation infrastructure, the CPD app's credentials are then revoked, and the users are notified.
Furthermore, due to the wide access to data given to the CPD apps, they are required to follow a strict security protocol in order for the CPD app to regain its credentials.

\subsubsection{Fourth Scenario: Stolen Credentials}
\label{sec:case:scenario-4}

The last scenario deals with stolen credentials, where an unauthorised subject managed to impersonate a legitimate subject to access the system.
Depending on the legitimate subject's credentials, such a scenario can threaten any or all of the CIA goals.

The specific authorisation goals for this scenario are the following:
\begin{itemize}
\item the authorisation infrastructure must detect the impersonation of legitimate credentials;
\item the authorisation infrastructure must ban subjects impersonating legitimate users, and force the victims to reset their credentials securely.
\end{itemize}

In this scenario, a malicious subject manages to steal the couple's credentials, and uses them to access the CPD app.
The authorisation infrastructure detects that, while the couple is currently located on one continent, a new connection seems to originate from a different continent. 
The authorisation infrastructure then automatically shuts down the malicious subject's access, as well as the access for the legitimate subject whose credentials were stolen. 
The legitimate subject then has to reset their password from an administration console provided by the EMS, which is only available from the household's local network.

\section{Related Work}
\label{sec:relatedWork}



In this section, we review some related work in the context of detection and mitigation of insider threat. 
Specifically, we discuss current approaches from three different solution areas.
These being dynamic access control, intrusion detection, and self-protection.
As such, it is intended to demonstrate the benefits and limitations of each solution area in mitigating insider threat, whilst arguing self-adaptation as a promising approach.

%
%


\subsection{Mitigation through Dynamic Access Control}
\label{mitigatingAtAccess}



Approaches to dynamic access control~\cite{Bistarelli:2010,Bohm:2010,janicke2012,Park:2004} are viewed as solutions to mitigating insider threat, due to their ability to enforce appropriate security controls given the state of the environment.
Observation of changes in environment state, such as, a rise in threat to national security~\cite{mcgraw:2009}, a dynamic access control approach will select an appropriate security control from a pre-defined set of controls, in order to mitigate attacks.
In the following, we discuss several notable approaches in dynamic access control, and their ability to mitigate insider attacks.


Usage Control (UCON)~\cite{Park:2004} builds upon traditional access control models whereby obligations and conditions are used to assess a subject's usage of a resource, as part of an access decision.
A novel aspect of UCON is its ability to capture a subject's state within a resource, and use this as a contributing factor within the access decision.
Whilst the UCON model is sophisticated in identifying and managing a subject's usage, it only allows for a transient solution to managing insider threat.
For example, a subject could invalidate usage requirements for a particular resource, but go on to access other resources despite being seen as a threat.
In addition, the UCON approach to access control has the potential to become complex because usage rules woven with traditional access control rules on a per resource basis.

A step forward from usage is the inclusion of trust and reputation when generating an access control decision, via a Trust Policy Decision Point~\cite{Bohm:2010}.
Here, a weighting of trust is calculated based on the usage or feedback from resources, providing additional context to a subject's usage.
Serrano et al.~\cite{Serrano:2009} explores trust management to achieve access control.
Within trust management, subjects and protected resources are given a level of trust, calculated from dimensions, such as, past behaviour of the subject, the access rights they already own, the issuer of access rights, and feedback from other subjects/resource owners.

In a similar work by Bistarelli et al.~\cite{Bistarelli:2010}, a formal framework for trust policy negotiation is proposed.
In contrast to Serrano et al.~\cite{Serrano:2009}, access is awarded through the reasoning of access control policies, and a trust level generated from a subject's given set of credentials.
An interesting aspect of Bistarelli et al. work is that not all subjects will know the required credentials for access.
Therefore, they propose an additional control that notifies the subject of the required credentials, providing the subject is deemed trustworthy. This adds an extra level of security, preventing the access requirements from being revealed unnecessarily, as they could be abused by a malicious subject.


There are other dynamic approaches specialised in expressing access control rules with a set of temporal constraints.
For instance~\cite{janicke2012}, access control policies contain a set of branch like rules, which are relevant to a set of system states.
Given a state that conforms to a temporal constraint or one that exhibits a particular event, access control mechanisms are constrained to a branch of relevant access control rules.
This approach to enabling dynamic access control (along with the aforementioned) is defined as \emph{dynamic policies}.

In summary, approaches to dynamic access control are capable of mitigating insider attacks, which is achieved through actively selecting appropriate security controls depending on the environment.
However, these approaches share a common limitation: it is necessary to maintain a comprehensive set of security controls in order to accommodate all potential risks of abuse.
As such, dynamic access control is seen as an improvement on traditional approaches to access control, but lack robustness regarding deployments that may fail to offer the necessary (and appropriate) security control, in light of attack.
In addition, dynamic access control requires fine grained security controls prescribed to particular states, which may prevent legitimate users from gaining access (due to constraints over time of access, or location) in order to prevent the prospect of malicious behaviour.
Finally, given that security controls are bounded to particular states (e.g., time of day), approaches are open to potential subterfuge where prevention of access is viewed as a transitive measure that could be overcome.
For example, if a subject abuses their credentials, from the viewpoint of a system administrator, one would expect that malicious subject's ability to access is removed entirely, whereas a dynamic access control approach may only temporarily prevent access (e.g., due to the time of day).

\subsection{Mitigation through Intruder Detection, Response, and Prevention}
\label{mitigatingAtNetworkLayer}

Intrusion detection systems are an established method of identifying and alerting system administrators to anomalous and malicious behaviour within a network.
 For detecting anomalous activity, they use a mixture of signature based rules based on known patterns of malicious packets over a network~\cite{Roesch:1999}, and machine learning techniques~\cite{Wang:2004}.
Whilst typically positioned for the detection of external attackers, recent works have demonstrated their use in detecting anomalous activity by malicious insiders~\cite{Bertino:2005}.

Intrusion detection alone is limited by a strong reliance on human administrators interpreting alerts and actively responding to alerts in order to mitigate insider attacks. 
Several works aim to improve upon this limitation through automated response to attacks, and the ability to prevent attacks from even happening. 
These are known as intrusion response systems (IRSs), and intrusion prevention systems (IPSs), respectively.

Intrusion response systems (IRSs)~\cite{carver:2001} work alongside intrusion detection systems to automatically respond to raised alerts.
Given the identification of certain types of attacks or alerts, an IRS will select a pre-determined response in order to mitigate any potential attacks.
Many of these approaches rely on a static decision making approach, such as Mu et al.'s~\cite{mu:2010} approach based on a hierarchical planning of responses (adaptations).
IRSs are capable in mitigating many forms of network based attacks, through both structural and parametric adaptation.
Example adaptations include the reconfiguration of network devices, adaptation of firewall policies, and throttling bandwidth to networked devices.

Intrusion prevention systems (IPSs)~\cite{Fuchsberger:2005} build upon IRSs, but sit within a network.
Rather than act on alerts, they actively monitor network traffic and perform adaptations as attacks occur.
IPSs have several advantages over IRSs.
Specifically, they relate to the timeliness in mitigating attacks without relying on the input from external systems (i.e., IDSs).
This in turn widens the scope of adaptations an IPS can perform over an IRS.
For example, an IPS is capable of immediately mitigating network based attacks through dropping or altering malicious packets in transit, or resetting network connections.

The advantage of an IRS and IPS solution is the ability to respond and prevent unauthorised and external attacks, where many adaptations involve changes to architecture (structural) or firewall rules (parametric).
IRSs and IPSs share similarities with self-adaptive approaches, yet are not explicitly classed as self-adaptive.
A key distinction to this is the lack of reasoning about `self', where many IRSs respond to alerts without considering the current state of the system and its environment.
Whilst both approaches are capable in mitigating attacks via adaptation, it is important for such systems to maintain an awareness of system state before and after adaptation. 
This would allow for the selection of optimal adaptations that can be evaluated against the current state, whilst providing assurances against adaptations that may cause greater damage to an organisation as opposed to allowing an attack to continue.

With respect to the mitigation of insider attacks, whilst IRSs and IPSs are well positioned to mitigating both internal and external behaviour, they are limited in mitigating attacks only at the network layer.
Internal attacks prolific of malicious insiders arguably offer different traits to that of external attackers intruding into a network, where there is a greater challenge in understanding the context of an insider's activity within a network, as well as their activity beyond the network layer.

Finally, one aspect of IRSs and IPSs that can benefit  a self-adaptive approach is the use of dynamic decision making in selecting responses (adaptations).
For example, some IRSs and IPSs make use of dynamic decision making in selecting adaptations to mitigate attacks.
Stakhanova et al.~\cite{stakhanova:2007} propose a dynamic decision making approach for IRSs where an IRS will analyse the results of a response to raised alerts.
The success or failure of a response is then factored into future decisions.
Ultimately, this can be extended to consider potential changes in perception of malicious or anomalous behaviour.

\subsection{Mitigation through Self-protection}
\label{sec:eng:adapt-auth}

Self-protecting systems are a specialisation of self-adaptive systems with a goal to mitigating malicious behaviour.
In the following, we discuss the few works that have demonstrated self-protection within the context of mitigating insider attacks.
In particular, we discuss two self-protection approaches based on the state of access control, and one approach based on the state system architecture. 

One of the approaches to self-protection via access control is SecuriTAS~\cite{Pasquale:2012}.
SecuriTAS is a tool that enables dynamic decisions in awarding access, which is based on a perceived state of the system and its environment.
SecuriTAS is similar to dynamic access control approaches, such as RADac~\cite{mcgraw:2009}, in that it has a notion of risk (threat) to resources, and changes in threat leads to a change in access control decisions.
However, it furthers the concepts in RADac to include the notion of utility, whereby given a perceived state of the system and its environment, the optimum set of security controls are used.
This is achieved through an autonomic controller that updates and analyses a set of models (that define system objectives and vulnerabilities, threats to the system, and importance of resources in terms of a cost value) at run-time.
The autonomic controller deploys optimal security controls (i.e., access control constraints) within the system, changing the conditions of access.
A novel aspect of this work is that it is aimed towards physical security, whereby a resource (e.g., a computer terminal or hand held device) is stored within an office (also considered to be a resource), for example.
SecuriTAS may change the conditions of access to the office based on the presence of high cost resources, or the presence of highly authorised staff.



Another form of self-protection in access control is positioned by SAAF~\cite{Bailey:2014}, a Self-adaptive Authorisation Framework. 
SAAF's goal is to make existing authorisation infrastructures self-adaptable, where an organisation can benefit from the properties of dynamic access control without the need to adopt new access control models.
This is achieved through a globally centralised autonomic controller that monitors the distributed services of an authorisation infrastructure to build a modelled state of access at run-time (i.e., deployed access control rules, assigned subject privileges, and protected resources).
Malicious user behaviour observed by a SAAF controller is mitigated through the generation and deployment of authorisation policies at run-time, preventing any identified abuse from continuing.
Adaptation at the model layer enables assurances and verification that abuse can no longer continue.
In addition, model transformation has been shown to generate authorisation policies from an abstract model of access.
This has the potential to enable the generation of policies specific to many different implementations of access control.

The main difference between SecuriTAS and SAAF, is that SecuriTAS positions its own bespoke access control model and authorisation infrastructure that incorporates self-adaptation by design. 
SAAF, on the other hand, is a framework that describes how existing access control models and authorisation infrastructures can be made self-adaptive, and as such, configured to actively mitigate insider threat.
With that said, both approaches demonstrate an authorisation infrastructure's robustness in mitigating insider attacks, by ensuring that authorisation remains relevant to system and environment states (and preventing continuation of attacks by adaptation of security controls).

In contrast to self-protection via access control, architectural-based self-protection (ABSP)~\cite{yuan:2013} presents a general solution to detection and mitigation of security threats, via run-time structural adaptation.
Rather than reason at the contextual layer of `access control',  ABSP utilises an architectural model of the running system to identify the extent of impact of identified attacks.
Once attacks or security threats have been assessed, a self-adaptive architectural manager (Rainbow ~\cite{Garlan:2004}) is used to perform adaptations to mitigate the attack.
One adaptation example  the approach offers is to throttle network connections to a server, in order to disrupt ongoing attacks.
Another example is the deployment of \emph{application guards} where a protective wrapper is deployed around architectural components (e.g., a web server).
These provide mitigation measures that improve upon the integrity of architectural components (i.e., the encryption of session ids susceptible to hijacking).
ABSP shares a number of similarities with intrusion response and prevention systems, particularly with the scope of adaptations that ABSP can perform (e.g., structural adaptation against network devices and connections).
However, because ABSP maintains a notion of `self', it is able to reason about the impact of adaptations and provide assurance over adaptation before adapting its target system.
\section{Self-Adaptive Authorisation Infrastructures}
\label{sec:dynamicAC}

In this section, in order to provide a basis for engineering
self-adaptive authorisation infrastructures, we map our perception of
self-adaptive authorisation infrastructures to the modelling dimensions for
self-adaptive software systems, as described by Andersson et
al.~\cite{Andersson:2009:MDS:1573856.1573859}.
To clarify this mapping, examples will be taken from the case study's scenarios introduced in Section~\ref{sec:case-study}.
A summary of the mapping is presented in Table~\ref{tbl:dimensions}.

\begin{table}
\centering
\caption{Summary of modelling dimensions for self-adaptive authorisation infrastructures}
\label{tbl:dimensions}
\begin{tabular}{ll}
\toprule
\textbf{Dimension} & \textbf{Degree}\tabularnewline
\midrule
\multicolumn{2}{c}{\textbf{Goals}}\tabularnewline
evolution & dynamic (main goals) or specific (specific goals)\tabularnewline
flexibility & rigid (main goals) or unconstrained (specific goals) \tabularnewline
duration & persistent (main goals), persistent temporary (specific goals)\tabularnewline
multiplicity & multiple\tabularnewline
dependency & independent or dependent \tabularnewline
\midrule
\multicolumn{2}{c}{\textbf{Change}}\tabularnewline
source & internal or external\tabularnewline
type & functional, non-functional or technological\tabularnewline
frequency & rare or frequent\tabularnewline
anticipation & foreseen, foreseeable or unforeseen\tabularnewline
\midrule
\multicolumn{2}{c}{\textbf{Mechanisms}}\tabularnewline
type & parametric\tabularnewline
autonomy & mostly autonomous, but sometimes assisted\tabularnewline
organisation & centralised\tabularnewline
scope & local or global\tabularnewline
duration & short, medium or long\tabularnewline
timeliness & best effort\tabularnewline
triggering & event-triggered \tabularnewline
\midrule
\multicolumn{2}{c}{\textbf{Effects}}\tabularnewline
criticality & harmless to mission-critical\tabularnewline
predictability & deterministic \tabularnewline
overhead & insignificant to failure\tabularnewline
resilience & ideally resilient, but this is hard to achieve\tabularnewline
\bottomrule
\end{tabular}
\end{table}

\subsection{Goals}
\label{sec:class:goals}

\subsubsection{Evolution}

The main goals of the authorisation infrastructure, e.g., confidentiality,
integrity, and availability, are inclined to be static, though they may change during the infrastructure lifetime.
This is the case when tradeoffs between goals need to resolved, which may require the renegotiation of some of the main goals.
However, specific goals may be more dynamic.
For example, the goals in the first scenario
(Section~\ref{sec:case:scenario-1}) pertain to protecting the infrastructure
against misuse by the third party service. 
These goals are only valid from the moment the household subscribes to the service, until they end their subscription.

\subsubsection{Flexibility}

Some of the main goals are rigid, as they prescribe that the infrastructure must
preserve confidentiality and integrity, for example. 
However, some specific goals may be constrained or unconstrained. 
The third scenario
(Section~\ref{sec:case:scenario-3}) provides a constrained goal, which
is that the infrastructure should detect and act upon when the integrity of the data is compromised. 
Unconstrained goal are not fixed: they may be
defined on a range of acceptable values, or as a particular situation
that must be handled, but without specifying how it should be
handled. 
The fourth scenario (Section~\ref{sec:case:scenario-4})
provides an example of an unconstrained goal, where the availability
of the services that are neither mission- nor safety-critical should
only be guaranteed according to a best-effort strategy. 
The goal does
not define a minimal acceptable availability value, nor does it
explicitly state \emph{how} the goal should be satisfied.

\subsubsection{Duration}

The main goals are persistent, as no breach in  integrity or confidentiality should be tolerated. 
However, once again, specific goals may be either persistent or temporary. 
The first
scenario provides temporary goals, in the sense that they are only
valid whilst the service has access to the system, which can be
revoked or granted at any time.

\subsubsection{Multiplicity}

The system has multiple goals. 
There are the three main goals,
availability, integrity, and confidentiality, which are common to all
self-adaptive authorisation infrastructures.
There are also the specific goals
described in the case study scenarios, that are specific to each
system or each deployment.

\subsubsection{Dependency}

The dependency between goals vary. 
For example, the confidentiality and integrity goals are independent. 
However, there is a dependency
between the confidentiality and availability goals, as well as between
the integrity and availability goals. 
A system enforcing maximum
availability may permit all requests, which would harm both the
confidentiality and integrity goals. 
The goals in the second scenario (Section~\ref{sec:case:scenario-2}) mandate that the system must be protected against misuse by a third
party service. 
These goals are complementary, and they also complement
the confidentiality goal. 

\subsection{Change}
\label{sec:class:change}

\subsubsection{Source}

The source of change in self-adaptive authorisation infrastructures can be external, internal, or both. 
A self-adaptive authorisation infrastructure that uses intrusion detection techniques will monitor external changes, while a system that monitors users for patterns of
misbehaviour will monitor internal changes. 
In the fourth scenario (Section~\ref{sec:case:scenario-4}), the detection that the geographical location from which request is made to access the CPD is different from the location in which the couple is known to be will be an external change. 
On the other hand, in the third scenario
(Section~\ref{sec:case:scenario-3}), the detection of suspicious
behaviour by the authorised CPD user will be triggered by internal
changes, because it is likely the access logs on the EMS that will be
monitored.

\subsubsection{Type}

Non-functional changes can be exemplified by the case of policy updates, such as the removal of the energy saving service's
credentials when a confidentiality breach is detected in the first scenario.  
Functional changes can also happen when a new goal is incorporated into the system, for example when a new service is
connected to the system, as in the first scenario.  
Finally, technological changes can also happen, if the system is able to change its policy language, or the policy evaluation engine, for example, when a security notice is issued for the running software.

\subsubsection{Frequency}

The frequency of changes also varies widely depending on the type of
change, and depending on the threats that the self-adaptive
infrastructure needs to protect itself against. 
Changes such as new
entries added to the authorisation log, that can be used in the first,
second, and third scenarios, are frequent. 
Other changes, such as the stolen credentials of the fourth scenario, are likely to be less  frequent. 
Finally, in the first scenario, the subscription to, or unsubscription from, the service, is a rare occurrence.

\subsubsection{Anticipation}

The degree of anticipation can also vary. 
There will certainly be foreseen changes, and the second scenario provides a good example since any system connected to the internet can reasonably expect automated attacks to happen quickly.
The system should be able to anticipate some of these attacks.
Foreseeable examples can also be captured by the second scenario: many variants of attacks will eventually happen, and it is not possible to foresee each and every variation of them. 
However, classes of attacks can be recognised, and attacks that fall into such classes can be detected and dealt with, even if they were not specifically foreseen. 
Finally, there may be unforeseen changes, i.e., changes that the system has not been designed to handle. 
Those could be dealt with using automated improvement or modification of existing detection and response mechanisms (e.g. using genetic algorithms), or sometimes through chance alone.

\subsection{Mechanisms}
\label{sec:class:mechanisms}

\subsubsection{Type}

The type of mechanism regarding self-adaptive authorisation is parametric. 
It is the authorisation policy that is changed by the self-adaptive authorisation infrastructure. 
The authorisation policy is specific to each deployment of the software: it contains the parameters that determine who can access what, under which circumstances.
Regarding self-adaptive access control, the type of mechanisms is parametric because self-adaptive authorisation infrastructure can change how an access control model is implemented, which implies manipulating the deployment of the different services of a federated environment (Section~\ref{sec:authinfrastructures}).

\subsubsection{Autonomy}

The self-adaptation mechanisms are preferably autonomous in self-adaptive authorisation, specially, if a potential intrusion is
detected. 
The credentials should be updated quickly in order to satisfy the confidentiality goal. 
The fourth scenario, in particular, require an autonomous mechanism. 
Since attacks may happen at all times, waiting for the user's input may take too long. 
The first scenario, however, may be implemented using both an autonomous or an assisted mechanism. 
If an assisted mechanism is chosen, the users may be notified that the service's pattern of requests is unusual, and they can decide whether to unsubscribe from it or not.

\subsubsection{Organisation}

The adaptation is centralised since there is a single component controlling all authorisation related services.

\subsubsection{Scope}

The scope of adaptation can vary between local and global, depending on the change made. 
A change that affects a user's ability to use a service is global when all the users are not able to use that service. 
The first scenario is such an example where the detection of suspicious behaviour of the application service can lead to the service's credentials to be entirely revoked. 
However, a local change could only affect a user's credentials for a specific service, without preventing the user from using other services.

\subsubsection{Duration}

The duration of a change may vary from short to long. 
A subject could be barred from using the system for a short period of time, up to permanently. 
The fourth scenario provides a example of a short change duration: once the external attack is detected, the system can change  to a more restricted access mode for a few minutes to a few hours, until the attack stops. 
The first scenario provides an example of a long change duration: once the application service subscription is cancelled and its credentials revoked, the change is permanent, unless
a human intervention subscribes to the service again.

\subsubsection{Timeliness}

The timeliness of changes is best-effort. 
It is difficult to offer guarantees when it comes to modifying the authorisation policy because any change may require extensive analysis. 
If many changes happen in a short period of time, it is unrealistic to hold any requests until the analysis has been done. 
Furthermore, all of our scenarios require the analysis of patterns of access, which requires the analysis to involve a number of changes.

\subsubsection{Triggering}

The changes that initiate adaptation are always event-triggered. 
The authorisation policy changes are made in response to detected events, which can happen at any time.

\subsection{Effects}
\label{sec:class:effects}

\subsubsection{Criticality}

The criticality of a failed self-adaptation may range from harmless to
mission-critical. 
If the self-adaptation fails, then the system will still be vulnerable to the threat it was trying to protect itself
against. 
The criticality of such a situation depends on the threat
itself, and whether unauthorised access can indeed be obtained. 
In the third scenario, an external adversary may be able to penetrate the system and turn it off in which case the self-adaptation failure would have mission-critical consequences. 
But if the adversary
does not manage to get into the system in the fourth scenario, then
the failure is harmless.

\subsubsection{Predictability}

The consequences of the self-adaptation are deterministic. 
It is always possible to find out what are the consequence in adapting a policy.

\subsubsection{Overheads}

The overheads caused by self-adaptation on the system's performance can
also vary, from insignificant to system failure. 
This depends on the
implementation of the system, as well as on the number of attacks
detected by the system.

\subsubsection{Resilience}

A self-adaptive authorisation infrastructure should be resilient. 
It is important to make sure that any adaptation to a policy, devised to resist a particular attack, will not make the system more vulnerable. 
Evidence should be provided that the self-adaptive authorisation infrastructure makes sound decisions that do not undermine the system resilience properties.

\section{Challenges in Engineering Self-adaptive Authorisation Infrastructures}
\label{sec:techniques}


In this section, we identify some challenges for engineering self-adaptive authorisation infrastructures
in the context of the MAPE-K loop.  
Specifically, for each of the
stages of the MAPE-K loop, we discuss what are the challenges
specifically associated with self-adaptive authorisation
infrastructures, looking, in particular, into issues related to
insider threats.  For example, what type of probes are needed for the
Monitor stage, how to generate dynamic plans in the Plan stage, and
how to perform policy updates in Execution, etc.  For each of the
challenges, we identify and describe the challenge, discuss their
relevance in the context of authorisation infrastructures regarding
insider threats, and provide an example related to the scenarios
identified for the case study previously defined (see
Section~\ref{sec:case-study}).


\subsection{Monitor}
\label{sec:monitoring}

The size of what needs to be monitored, and the ability of the
monitoring to adapt its own probes and gauges are the two dimensions
that will influence the complexity of the Monitor stage.
Since self-adaptive authorisation infrastructures have no control over their
environment, it is impossible to foresee \emph{all} the environment
changes that might affect the system.
Some changes can easily be
detected by the probes and gauges of the self-adaptive authorisation infrastructure, while some others can remain oblivious if
the appropriate probes and gauges are not provided.
In order to avoid
the risk of the infrastructure missing important information, it is
necessary to dynamically adapt (1) what needs to be monitored, and (2)
the type of probes and gauges required.




\subsubsection{Active Monitoring}
\label{sec:eng:monitoring:active}

With passive monitoring, static probes and gauges are set up at
deployment time, to monitor the authorisation infrastructure
and its environment.
The probes and gauges are static since they
cannot be re-deployed or removed at run-time, nor can they be
re-configured.

While it may be tempting to monitor a wide range of environment resources, monitoring comes at a cost.
It has an impact on performance, and may affect other requirements, such as the users' privacy.
Within a changing self-adaptive authorisation infrastructure and its environment, the right balance between data collection and performance or privacy is likely to evolve.

\paragraph{\textbf{Challenge.}}

The challenge is the provision of active (or pro-active) monitoring for reducing the amount of traffic related to monitoring considering that some of the analysis can be performed by the probes and gauges themselves, without sending the data to the controller for analysis.
Moreover, pro-active monitoring requires the availability of smart probes and gauges, able to adapt to what they monitor.



\paragraph{\textbf{Relevance.}}

The key motivation for pro-active monitoring in self-adaptive
authorisation infrastructures is to make dynamic access control more
resilient to changes, thereby allowing the infrastructure to better
detect and react to insider threats.
The detection of insider threats
relies on monitoring a wide range of resources from the environment of
the authorisation service with the purpose of profiling the status and
activity of subjects inside the organisation.
As the monitoring might be outside the ownership of the authorisation service, special probes need to be synthesised and deployed that might be constrained by privacy issues, for instance.

\paragraph{\textbf{Example.}}

In the first scenario, the couple may fall into a routine, leaving and coming home around the same time every day.
An active gauge is monitoring the users' location, to turn the heating on and off.
However, privacy considerations require to keep this to a minimum.
The gauge identifies a pattern in the users' location, and may choose to only query the users' smartphones around the time where it expects a change in location.
Upon detecting an unusual location, the gauge itself decides to increase its monitoring frequency, in order to feed the system with more data to detect a potential insider threat.

\subsubsection{Run-time Synthesis of Probes and Gauges}
\label{sec:eng:monitoring:runtime-synthesis}


The synthesis of probes and gauges at run-time is one way of achieving
active monitoring.
While the decision to synthesise probes and gauges
may be out of the scope of the Monitor component of the MAPE-K loop,
their synthesis, configuration, and deployment, are not.

\paragraph{\textbf{Challenge.}}

The challenge is the ability to synthesise probes and gauges at run-time, in response to new or emerging attacks.
These probes and gauges, once deployed, should improve the resilience of the self-adaptive authorisation infrastructure against unexpected changes.

\paragraph{\textbf{Relevance.}}

The run-time synthesis and deployment of probes and gauges in self-adaptive authorisation infrastructures can help to cope with the unpredictable nature of an attack.
There are no guarantees that what is being monitored is sufficient to identify a whole range of attacks,
hence the need to autonomously synthesise and deploy a probe or a gauge that would be able to examine novel system attributes.

\paragraph{\textbf{Example.}}

The first scenario in the case study is a good illustration of this:
when the users subscribe to the third party service, a new subject
gets access to the system, thus this new subject must be monitored.
A system that is not very resilient will rely on the user to create, configure and deploy a new set of probes and gauges to monitor the
behaviour of the third party service.
On the other hand, a resilient system should be able to extract a set of probes or gauges from a repository, before
configuring and deploying them automatically.
However, a more resilient system should be able to create a new set of probes and gauges
tailored to the service to be monitored, and configure and deploy them
automatically.

\subsubsection{Mutating Gauges}
\label{sec:eng:monitoring:mutating}

Mutating gauges are gauges that are able to change themselves, either randomly or guided, in order to identify unknown behavioural patterns that might be related to an attack.
If the monitoring system needs to have the capability to detect autonomously previously unknown patterns of attack, one way to enable this is to generate new detectors by mutating existing ones.

\paragraph{\textbf{Challenge.}}

The challenge is to generate and deploy these mutating gauges for examining real-time or past data to identify unexpected interactions that an authorised subject might have with the system being protected.
These mutating gauges can be used to provide additional evidence, with some degree of confidence, that an attack is, or has been, taking place.

\paragraph{\textbf{Relevance.}}

Since the environment of authorisation infrastructures are dynamic and unpredictable, one should not expect to know about all possible attacks before deploying the system.
The ability to deploy mutating gauges would enable the detection of new forms of attack by simply looking for unknown anomalies, and this would be enabled by the random nature of these gauges, i.e., there is no implicit expectation of what they should be able to detect.
Lets consider the case in which a gauge monitors the access to a service by authorised users.
A possible change in the environment of this service is the deployment of a new version of the server providing the service, and this might result in changing the format of the logs that the gauge is supposed to monitor.
Either the original gauge becomes ineffective, or it needs to be manually re-configured.
Alternatively, once a change is detected in the log format, a mutating gauge may be able to automatically adapt itself in order to understand the new format.
Another possible usage
of mutating gauges would be to enable the perpetual analysis of logs in order to identify attacks.
During run-time, as an offline activity, different gauges could be dynamically generated by mutation, and these would analyse the logs for identifying attacks previously unknown.

\paragraph{\textbf{Example.}}

The first scenario of the case study provides an appropriate context
in which mutating gauges might be useful.
A simple gauge could monitor the third party service when accessing the couple's location by
triggering an alert only if the location is accessed more than once
during a specified time frame.
Mutations of the gauge could record
more complex data about the third party service's queries.
For example, a mutated gauge could keep track of the frequency of queries
over time, and trigger an alert if it suddenly increases, which would
provide the couple with a more precise way of finding out that their
privacy is likely to be under threat.

\subsubsection{Incomplete Information}
\label{sec:eng:monitoring:incomplete}

Incomplete information refers to the situation in which the Monitor stage is not able to provide all the information needed by the other stages of the control loop.
This might be due to limited monitoring capabilities, and because of that, the monitoring stage has to find alternative ways of obtaining the missing information.


\paragraph{\textbf{Challenge.}}

Identify and select what to monitor in order to compensate the missing information, and know where it is safe to make assumptions about the unknown.

\paragraph{\textbf{Relevance.}}

Monitoring has a cost, especially when considering insider threats.
The detection of insider threats relies mostly on data from the environment, and since the environment of an authorisation infrastructure is broad and fluid, in the sense that it is difficult to establish its clear boundaries, this has an effect on the data that is collected.
Therefore, the system will likely have to deal with incomplete information, which in the Analyse stage might lead to more false positives regarding insider threats.
One way to compensate for incomplete information is for the gauges themselves to provide a level of confidence regarding the information that is forwarded to other stages of the control loop.

\paragraph{\textbf{Example.}}

The fourth scenario of the case study could benefit from
such a feature in which a confidence level could be incorporated into
the monitored information.
The location of a user can be determined via several techniques, such as the IP address, triangulation from mobile network towers, or GPS signal.
These techniques have varying levels of precision and reliability.
Furthermore, a malicious user may also tamper with the readings in
order to fool the system.
If a gauge is able to attach a measure of confidence (obtained, for example, through several probes using different techniques to capture the user's location), then it would help the other stages of the control loop in deciding whether the user credentials have likely been stolen or not.

\subsubsection{Automatic Feature Identification}
\label{sec:eng:monitoring:feature-ident}

During system operation certain probes may cease to function, either
maliciously or accidentally.
In order not to lose the features being
monitored through that probes, the system should be able to recover
some or all of those features by making use of the information
provided by other probes.
The assumption is that several features
can be associated with a probe, and that these features can be
extracted and combined with other features from other probes in order
to reconstruct totally or partially the information lost from an
unavailable probe.

\paragraph{\textbf{Challenge.}}

The challenge is for the system to be able to automatically extract features from its probes, and recombine those features as necessary, thus exploiting some intrinsic redundancy that may exist amongst the probes.
At run-time, this should be achieved by combining and reconfiguring features that are associated with the information provided by several probes.

\paragraph{\textbf{Relevance.}}

This challenge is relevant to self-adaptive authorisation infrastructures because it
helps to increase the system's resilience against run-time threats to probes and gauges,
whether they are intentional or accidental.

\paragraph{\textbf{Example.}}

The fourth scenario illustrates an application for automatic feature identification.
In this scenario, if a denial of service attack is detected, the system can be reconfigured to ensure the availability of its critical services.
The system reconfiguration does not have to be limited to protecting itself from an attack since the re-combination
of the probes' features may also allow the system to reduce the monitoring overhead.
If probes themselves were affected by the denial of service attack, then recovering the affected probes' features using the information from other probes would allow the system to maintain the same level of monitoring.

\subsection{Analyse}
\label{sec:analysis}


The Analyse stage is made of two consecutive parts: the problem domain analysis and the solution domain analysis.
The problem domain analyses the data provided by the Monitor stage in order to identify changes that
the system may have to respond to.
The solution domain analysis occurs after a problem has been identified, and is concerned with generating possible
alternative solutions that are able to handle the problem.
The problem domain analysis can be further divided in two parts:
the identification of potential problems, and the assessment of the
identified problem in order to prioritise the mitigation.
In some cases, a problem may be identified as sufficiently serious to be
addressed immediately.
At the other end of the spectrum, some
problems may be acknowledged, but ignored, as they are deemed not
critical enough to cause adaptation.

In the following, first, we present the challenges related to the
problem domain: anomaly detection, signature-based detection,
case-based detection, diagnosis, and normality detection.
Then, we
present those challenges related to both problem and solution domains
by clearly identifying how these are related to each of the domains:
perpetual evaluation, threat management, and risk analysis.

\subsubsection{Anomaly Detection}
\label{sec:eng:analysis:anomaly-detection}

Anomaly detection is
related to the ability of the controller to identify any behaviour
that deviates from what is perceived to be acceptable.
Since we are essentially dealing with socio-technical systems for which it is
almost impossible to establish, from the outset, all their possible
behaviours, it is extremely challenging to clearly distinguish normal
from abnormal behaviour, i.e., what is acceptable and what it is not.
First, there is the uncertainty of the context of the system that
might influence whether a particular behaviour is deemed to be normal
or abnormal.
Second, there are the previously unknown or unexpected
behaviours that need to be classified according to profiles of similar
class of behaviours.

Since  there is no single technique that should be able to
accurately detect a wide range of anomalies, one way of reducing the number of misclassifications is to use
diverse techniques whose outcome should be fused for providing confidence in the classification.
In the following, after introducing anomaly detection challenge, we present, as an example, two specific complementarity anomaly detection
techniques that can be used for improving both the responsiveness and coverage when detecting anomalies.

\paragraph{\textbf{Challenge.}}
The key challenge in anomaly detection is to be accurate when
detecting anomalies under uncertainty in order to reduce misclassifications, specifically in the context of insider threats.
Since misclassifications cannot be eliminated, it is important to associate with those classifications levels of uncertainty.

\paragraph{\textbf{Relevance.}}
The ability of detecting anomalies should precede the system
capability of handling insider threats.
Since it is difficult to
accurately identify an attack, uncertainty levels should be considered
so the system can evaluate a particular detection against its context.
The objective is to reduce the number of false positives and false negatives
that might have detrimental consequences upon self-adaptive
authorisation infrastructures.

\paragraph{\textbf{Example.}}
The first scenario, regarding reading and writing requests, motivates quite well the need for having accurate
detection of a misuse.
If a particular abuse is not detected in time, the privacy of the couple
might be compromised.
All the other scenarios also quite motivate the
need for having an effective and efficiency means for detecting
anomalies because the failure of not detecting an abuse might
compromise the whole system.

\subsubsection{Signature-based Detection}
\label{sec:eng:analysis:signature}

Signature-based detection is a special case of anomaly detection (see Section~\ref{sec:eng:analysis:anomaly-detection}), where domain analysis is performed by matching the data
provided by the Monitor stage against signatures of known problems.
A signature is a pattern that should be matched against the data
provided by the Monitor stage, such as, an IP address, a particular
regular expression in a log file, a URL, a version of some software,
etc.
Signature-based detection may require the matching of several
individual signatures to identify a threat.
They are relatively easy to automate.
Since signatures refer to precise
pieces of information, it is possible to completely automate their
recognition, and therefore the identification of threats.
With a sufficiently expressive language to write the
signatures and their interactions, complex analysis can be performed
to discover advanced threats.
Administrators should also be
allowed to define signatures, as well as, combinations of signatures,
and associate them with threats.

\paragraph{\textbf{Challenge.}}

Since the signature-based detection is a static technique, the challenge is to be able to synthesise new signature-based detectors during run-time.

\paragraph{\textbf{Relevance.}}
Signature-based detection is best suited to detect threats that are known and well understood in advance.
However, in the context of self-adaptive authorisation infrastructure the efficacy of static signature-based detectors is quite restrictive considering that both the attacks and the infrastructure can change.
Thus the need for the self-adaptive authorisation infrastructure to be able to generate dynamically new signature-based detectors that are able to detect unknown threats efficiently at run-time.

\paragraph{\textbf{Example.}}

The second scenario provides an example for a simple signature-based detection algorithm.
If the energy retailer is able to provide a wide range of services from different IP addresses, the self-adaptive authorisation infrastructure needs to identify those legitimate services  that might abuse their privileges.
For example, the authorisation infrastructure needs detects that a particular service from the
energy retailer, originating from a particular IP address, reads the couple's energy use too often, which may result in
a potential threat to the couple's privacy.
The detection of the
energy retailer's behaviour can be implemented using signature-based
detection, where the signature of the threat is a number of
connections from the energy retailer's IP address that exceeds a
pre-determined threshold in a pre-determined time frame.

\subsubsection{Case-based Detection}

Case-based detection is another special case of anomaly detection, where the focus is on observing subjects' behaviours, which are harder to model, and hence, harder to automate.
Where signature-based detection attempts to
identify well-defined actions performed by malicious subjects, case-based detection observes the malicious behaviour of subject, and allows
for decisions to be made based on the subject's behaviour model.
Moreover, instead of absolute thresholds for identifying anomaly detection, relative thresholds comparing users behaviours can be used.

\paragraph{\textbf{Challenge.}}
The challenge in case-based detection involves recognising a behaviour that may not be
explicitly forbidden, but still suspicious.

\paragraph{\textbf{Relevance.}}
It may be the case that a subject will try to circumvent signature-based detection since
signature-based detection works by using thresholds and precise patterns of attacks.
This is where case-based detection becomes useful.
The attacker may be slowed down because of their efforts
to avoid detection, but that does not mean that the threat does not need to be addressed.
Case-based detection is a good way to
complement signature-based detection because of its ability to detect
and act upon those types of threats, although it is more difficult to be fully automated.

\paragraph{\textbf{Example.}}
Since case-based detection can complement signature-based detection, we use the same example to illustrate both approaches.
With signature-based detection, the intrusive energy retailer was detected
when they read the couple's energy consumption more than a pre-defined
number of times during a given time frame.
One way for the energy retailer to avoid detection by the signature-based detection system is
to stay right under the threshold.
Finding out what the threshold may have involved getting caught once.
The case-based detection system may be looking at the history of the read operations by the energy
retailer.
This analysis may identify that the frequency went up until they got caught by the signature-based detection system, before
staying just under the threshold.
This may be constructed as suspicious behaviour, especially if the retailer had previously
performed much less read operations per time period.
Similarly, the detection of abuse could be related to a dynamic threshold instead of a static one.
By profiling the number of times the energy retailer
reads the energy consumption within particular time intervals, these can be compared for detecting an abuse.
An administrator may be
notified and shown a model of the retailer's behaviour to decide
whether is their can be characterised as intrusive

\subsubsection{Diagnosis}
\label{sec:eng:analysis:diagnosis}
When an attack is detected, the system may try to identify the source of the
attack, how it was performed, what damage it caused or is causing, and
which vulnerability was used to carry it out.
Diagnosing an attack allows the system to better understand it, and therefore to make better decisions to defend against it.

\paragraph{\textbf{Challenge.}}
The challenge of diagnosing self-adaptive authorisation infrastructures is the ever changing type of attack, and the new vulnerabilities that might be introduced during adaptation.

\paragraph{\textbf{Relevance.}}
Identifying the source of the attack and the vulnerability exploited is key to stopping it to propagate, as well as making sure
that it does not happen again.
A self-adaptive authorisation infrastructure that can understand where attacks come from and how they are carried out will
be more resilient than a system that can only identify them without understanding what caused them to be successful.

\paragraph{\textbf{Example.}}
The third scenario (Section~\ref{sec:case:scenario-3}) can be handled
much better by a self-adaptive system capable of performing diagnosis
on the attack.
A system that does \emph{not} attempt to understand
the attack may restrict its actions to modifying the user's credentials.
However,
if the attacker used a vulnerability in the system to gain the user's
credentials, they can run the attack again once the credentials have
been re-configured.
A system that would perform advanced diagnosis,
however, may be able to identify how the attacker got hold of the
credentials, and may be able to solve the root issue, or give the
administrator useful data for them to do so manually.

\subsubsection{Resuming Normality}
\label{sec:eng:analysis:normality-detection}
When an attack is over and a threat does not anymore pose danger, or
when a particular risk that had previously identified has been
mitigated, the system should be able to undo the restrictive measures that were
taken for protecting the system against the attack,  or the likelihood of an attack.
This would be more relevant if the restrictive measures taken affected the system's normal operation.
This should be done without exposing the system to other attacks.

\paragraph{\textbf{Challenge.}}
After taking measures to protect the system against attacks, the challenge is when to undo some of the restrictive measures, and what measures should be put in place in order maintain a balance between usability and security.

\paragraph{\textbf{Relevance.}}
Measures taken to prevent or mitigate attacks, in the context of
self-adaptive authorisation infrastructures, often take the form of
reduced capabilities for users, or more stringent authorisation
procedures.
If the system were not able to scale back some of the
measures taken after the event that triggered them has occurred, then
the system would tend towards locking all the users out of the system.
It is therefore crucial that the system is able to always strike the
correct balance between usability and security.

\paragraph{\textbf{Example.}}
The second scenario (Section~\ref{sec:case:scenario-2}), where the
intrusive energy retailer is prevented from reading energy consumption
data if they have done it too often, requires for the counter-measure
to be eventually lifted.
This can be done after the self-adaptive authorisation infrastructures  has assessed that the energy provider's behaviour does not pose
a threat anymore.
Failure to do so would prevent the provider from reading data that is essential to correctly billing the users.

\subsubsection{Perpetual Evaluation}
\label{sec:eng:analysis:perpetual-evaluation}

When the controller is not adapting the target system, it can run background tasks to enhance the resilience of the self-adaptive authorisation infrastructure.
Perpetual evaluation is one such task, which stands for the continuous analysis of either the problem or solution domains.

\paragraph{\textbf{Challenge.}}

The challenge associated with the perpetual evaluation of the problem domain  is the identification of vulnerabilities and attacks that might affect the  self-adaptive authorisation infrastructure.
On the other hand, the challenge associated with the perpetual evaluation of the solution domain is the provision of assurances regarding the quality of services provided by the self-adaptive authorisation infrastructure.


\paragraph{\textbf{Relevance.}}

Perpetual evaluation can be used alongside traditional evaluation in
order to improve the coverage in detecting insider attacks, localise vulnerabilities, and
enhancing the provision of assurances.
This can be done either proactively or reactively.

If insider attacks can be predicted to occur depending on some observable pattern of behaviour, adaptation can be proactive, and the same applies to evaluation.
Since the proactive perpetual evaluation does not block any immediate adaptation, it can only  inform future adaptations.
While traditional evaluation can make fast, but imperfect, decisions, the reactive perpetual evaluation compliments traditional evaluation by
confirming that the adaptation satisfies the system goal, or point to issues that may require a rollback, or further adaptation.
This is possible because the reactive perpetual evaluation can afford to take longer to
complete, and consider more data or more stringent constraints.
This is especially useful in scenarios where timeliness of adaptation is
important, such as the response to insider threats.


\paragraph{\textbf{Example.}}
The third scenario (Section~\ref{sec:case:scenario-3}) illustrates the
advantages of perpetual evaluation.
Detecting suspicious deletions
should be done very quickly, as one would like to minimise the data
loss.
Ideally, suspicious behaviours should be detected before any
data loss happens.
However, it may not be easy to tell the difference
between legitimate and illegitimate deletion.

Proactive perpetual evaluation could monitor the system's access logs, and
compare each user's actions to their behaviour profile, and the
behaviour profile of similar users.
If a user starts acting
suspiciously, adaptation could be triggered before the users starts
deleting sensitive data.

Using reactive perpetual evaluation could allow the system to suspend a user's
permissions when suspicious activity is detected, like the deletion of
a large number of files.
This can be achieved quickly using
traditional evaluation techniques.
The reactive perpetual evaluation could then consider
more elements, such as a longer history of the user's access data, or
access data for similar users, to determine with more confidence
whether the file deletions were likely to be legitimate or not.
If
they were legitimate, a plan to reactivate the user account can be
made.
If they were not, it confirms that the decision taken by the
traditional evaluation was the right one.

\subsubsection{Threat Management}
\label{sec:eng:analysis:potential-attacks-management}

There may be several simultaneous attacks detected or vulnerabilities identified, and responses to these in the form of adaptations should be prioritised.
Furthermore, responses may increase the attack surface, or weaken other security measures.

\paragraph{\textbf{Challenge.}}

In the problem domain, the ability to prioritise attacks  and vulnerabilities is a challenge associated with threat management, which should take into account the threats' potential impact on the system's operations, and attempt to take preventive measures, to ensure that future threats can be addressed.

In the solution domain, the challenge associated with threat management is the ability to rank alternative responses, and to ensure that a response does not increase the system's attack surface, or weakens its security measures.


\paragraph{\textbf{Relevance.}}
Any perceived attack or vulnerability should not be considered in isolation from its
current or historical contexts, otherwise problem domain analysis
might be incomplete, thus producing outcomes that might undermine the
mitigation of threats.
Likewise, from the solution domain
perspective, any measure to handle the perceived attack or vulnerability should take
into account other measures either being processed or already
processed.
The goal is to reduce the amount of resources needed for
handling the attack or the vulnerability, and minimise the risk of introducing new
vulnerabilities.
Moreover, considering that known vulnerabilities
might exist in the authorisation infrastructure, these should be taken
into account when analysing measures for mitigating a perceived
attack.

\paragraph{\textbf{Example.}}
The attack in the first scenario (Section~\ref{sec:case:scenario-1}) combined with the attack of the second scenario (Section~\ref{sec:case:scenario-2}).
In this situation, the self-adaptive authorisation infrastructure should decide which one of the attacks is or higher risk, respectively, the compromise of integrity in the services provided by the third party, or the privacy violation by energy retailer.


\subsubsection{Risk Analysis}
\label{sec:eng:analysis:risk-analysis}

When perceived to be under attack, an authorisation infrastructure can be used risk levels to rank alternative responses and select the most appropriate one.
Factors that can influence the risk level include the coverage of the evaluation, the severity of the attack and/or the vulnerability, but also the impact of countermeasures on the system's operations.

\paragraph{\textbf{Challenge.}}

The challenge of risk analysis in the problem domain is to determine the seriousness of an attack, which should establish the appropriate response level.
Regarding the solution domain, risk analysis should guide the selection of the most appropriate response when several options are available.


\paragraph{\textbf{Relevance.}}
In the problem domain, depending on the perceived risk, attacks and vulnerabilities may need to be
dealt with immediately, while others may allow for a delayed response, or no response at all, at little to no cost on the system's security.
Whether a self-adaptive authorisation infrastructure shall react to an
attack should depend on the risk associated with the attack and/or vulnerability:
the probability of an attack to be successful, and the impact the attack
might have on the system in case is not mitigated.

Regarding the solution domain, adaptation may be expensive, whether in terms of
time, computation resources, or inconvenience to legitimate users through
degradation of the service.
Therefore, a sophisticated authorisation
infrastructure could use risk analysis to prioritise the order in which
attacks should be addressed, and when and how to deal with them.

\paragraph{\textbf{Example.}}
A combination of the second and third scenarios
(Sections~\ref{sec:case:scenario-2} and \ref{sec:case:scenario-3},
respectively) illustrate the need for risk analysis.
In the second
scenario, the system detects that the energy retailer reads data from
the meter more often than agreed.
This issue needs to be addressed, but might not be the most critical.
The third scenario, however, deals with the detection of data being deleted by a malicious user.
This should be stopped as soon as possible since more data will be
lost until the issue is addressed.
If issues arising from both
scenarios are detected around the same time, and if the system is only
able to deal with one issue after the other, then the data deletion
attack should be addressed before the energy provider attack.

\subsection{Plan}
\label{sec:plan}

The Plan stage is made of two consecutive parts: decision making and
plan synthesis.
The purpose of decision making is to select the most
appropriate solution amongst the alternatives provided by the solution
domain analysis.
Below, we have identified three challenges
associated with decision making: decision making in a federated authorisation
infrastructure, randomising decisions, and denial of service.
The goal of plan synthesis is to generate a plan that implements the
selected solution.
We identified six challenges related to the plan
synthesis: robust plans, controller capabilities, and infrastructure boundary.

\subsubsection{Decision Making in a Federated Authorisation Infrastructure}
\label{sec:plan:decision:federated}

There are several benefits associated with federated authorisation infrastructures, being one of them the ability of authenticating users using third parties.
However, these pose additional challenges to the planning phase, compare with a simpler, centralised infrastructure over which a single entity or user has complete control.
The selection of the best solution amongst alternatives, identified during the analysis problem domain, needs to consider the self-interests of the different parties of the federation.
The component systems of a federated authorisation  infrastructure may have conflicting interests and goals, and varying constraints (e.g., an identity provider service may conflict with a service provider).
Yet it is important to be able to select the best solution amongst those  identified in the analysis solution domain, while satisfying the goals and constraints of all the components in the federated authorisation infrastructure.

\paragraph{\textbf{Challenge.}}

Decision making in a federated authorisation infrastructure should take into account the potentially conflicting goals of all the parties in the infrastructure, and negotiate a solution that satisfies them all.
This may require a solution that is not optimal, but ``good enough'', and acceptable to all parties involved.

\paragraph{\textbf{Relevance.}}

In a self-adaptive federated authorisation infrastructure, it is expected for third parties to undergo some kind of change, for example, involving their goals or their deployment.
This should have an impact on how the different parties collaborate in order to maximise each party self-interest.
However, adaptation decisions that involve federated authorisation infrastructures may require negotiation between several stakeholders.
If all the component systems' goals cannot be satisfied, then a self-adaptive authorisation infrastructure may have to consider stopping its collaboration with some or all of the component systems with whom a compromise could not be reached.

\paragraph{\textbf{Example.}}

In the fourth scenario, user credentials are stolen by a malicious
subject.  
If the resources under attack are protected by a federated
authorisation infrastructure where identity management is handled by
third parties, the self-adaptive system may have to negotiate with the
identity providers in order for them to take action against the
malicious subject.  However, should the third party fail to meet the
security expectations of the self-adaptive authorisation
infrastructure, the infrastructure may decide to revoke its trust in
the third party, and forbid all authentication tokens coming from it.

\subsubsection{Randomising Decisions}
\label{sec:plan:decision:randomised}

If the decision maker, for particular operational context, always
selects the same strategy, then a new vulnerability is being
introduced.
If an attacker, while interacting with the system or
observing its behaviour, is able to establish deterministically the
response of the self-adaptive authorisation infrastructure, the
attacker may be able to take advantage of this adaptation, and cause
harm to the system.

\paragraph{\textbf{Challenge.}}

The selection of an adaptation solution among several more-or-less equally acceptable options should be randomised, in order to prevent an attacker from learning about the system's response to a particular output, thus reducing the attack surface.


\paragraph{\textbf{Relevance.}}

Self-adaptive authorisation infrastructures could be targeted by attackers wishing to exploit a new vulnerability introduced by the controller.
The nature of the adaptation measures that can
be taken poses at least two threats.
First, the attackers could trick
the self-adaptive system into banning users, or groups of users, even
if only for a limited amount of time, causing disruptions in the
users' ability to use the service.
Second, the attackers could
trigger a denial of service attack by forcing very frequent changes in
the authorisation policies, which would overwhelm the system.

\paragraph{\textbf{Example.}}
In the first scenario, a compromised service gets access to the
system.
The compromised service could attempt to learn how the
authorisation infrastructure adapts by purposefully attempting to
trigger self-adaptation, and finding out how the system reacts to
certain events through observation.
If the response to specific
threats is not always the same, it will be much more difficult for the
compromised service to learn how the system works, and how it could be
disrupted.

\subsubsection{Denial of Service}
\label{sec:plan:decision:dos}

Denial-of-service (DoS) aims to make resource unavailable to their
legitimate users, for example, by flooding a server with bogus requests
that waste computing resources.
An attacker could use the self-adaptation mechanism for this purpose, preventing legitimate
users from using the service.

\paragraph{\textbf{Challenge.}}

As a challenge, the self-adaptive authorisation infrastructure should
be able to analyse the triggers for self-adaptation, identify their
source and frequency, and react accordingly in order to avoid the system to become unusable.

\paragraph{\textbf{Relevance.}}

Authorisation infrastructures for which the execution of self-adaptation requires reloading configuration
files, restarting services, or interrupting or cancelling long-lived
operations, are particularly vulnerable to DoS  attacks.
If the attacker finds a way to trigger self-adaptation often enough, the
system may become unusable for legitimate users.
In this case, the
self-adaptation mechanism itself is the attack vector used by the
attacker to perpetrate his attack.
If the system detects a possible DoS attack, it
may then switch to a less obtrusive means of self-adaptation if
available, or disable self-adaptation for some time.
Another option
would be to cap the number of self-adaptation operations that disturb
the service for a specified time period.

\paragraph{\textbf{Example.}}

The compromised service in the first example could use its credentials
to trigger a DoS attack.
It could find out which operations will likely trigger a
self-adaptation that will render the service unavailable for some
time, and then find a way to trigger that self-adaptation often enough
that the system will be unusable by legitimate users.
An attack
that would trigger a restart of the authentication server and the
revocation of user sessions would be a good example.
Users would have
to constantly re-authenticate and would not be able to use the system
properly.
If, however, the system were to be able to detect such an
attack, it could change its adaptation strategy so that legitimate
users are not affected, for example, by banning the compromised service
until it is manually re-activated by an administrator.

\subsubsection{Robust Plans}
\label{sec:plan:robust:plans}

Some of the activities in a plan may be more likely to fail than others.
This could be related to complex interactions between components of a federated authorisation infrastructure.
This can be caused by software or hardware failures, or simply because assumptions made during the conceptualisation of the plan cease to be true during its execution.

\paragraph{\textbf{Challenge.}}

In a self-adaptive authorisation infrastructure, the challenge is to obtain a robust plan that should be able to handle failures in one or several of its activities, while minimising service interruptions.
The plan should incorporate redundancy in its activities, or the ability to rollback to a previous working secure state, in case an activity fails.

\paragraph{\textbf{Relevance.}}

Authorisation infrastructures involve various component systems, as well as a number of policies whose interactions determine who gets access to what.
If a plan is not robust enough, the infrastructure could be left in an intermediate insecure and unstable state, i.e., some authorisation decisions could allow unauthorised subjects access to sensitive data.
If a plan is rolled back, then the infrastructure is again vulnerable to the insider threat that had triggered the (aborted) adaptation.
None of these scenarios are acceptable, and therefore, the plan should be as robust as possible in order to deal with any unexpected issue arising during its execution.
This can be achieved by enabling the controller to generate abstract plans (i.e., a plan that does not depend on any particular implementation, it can support several alternative implementations) that can be instantiated into concrete plans during their execution.
In case a particular instantiation of an activity fails during its execution, the abstract plan should incorporate enough redundancy in order to activate an alternative instantiation.

\paragraph{\textbf{Example.}}

The first scenario illustrates the need for a robust plan.
Once the compromised service has been detected, the self-adaptive authorisation
infrastructure should synthesise a plan that would remove the
service's credentials from the authorisation policy, and would reload
the new policy into the authorisation infrastructure.
If, for some
reason, the policy cannot be properly reloaded into the
infrastructure, then the system is left into an intermediate state in
which the compromised service still has access to the system.
A plan that incorporates redundancies could, for example, force the whole authorisation infrastructure to be restarted using a restricted but trusted policy, if a new valid policy cannot be reloaded.

\subsubsection{Controller Capabilities}
\label{sec:plan:synthesis:capabilities}

What a controller is able to achieve in a self-adaptive authorisation infrastructure is restricted by its capabilities.
These capabilities are related to what the controller is able to observe and control, and its computational and algorithm resources.
Limitations on the controller's capabilities might have an impact on the plans that a controller is able to synthesise, and these limitations should be incorporated into plans.

\paragraph{\textbf{Challenge.}}

In a self-adaptive authorisation infrastructure, because of its nature, it is difficult to forecast changes that  might affect the system and its environment, the challenge is for the controller to be able to identify its own limitations.
In case an operational boundary is reached, the controller should be able to act, either by shutting itself down or invoke another alternative controller, for example.

\paragraph{\textbf{Relevance.}}

While synthesising a plan, there is a risk that some implementations are not able to support activities of the plan.
For example, the controller of a self-adaptive authorisation infrastructure is able to synthesise plans that only contain activities that rely on XACML implementation of Policy Enforcement Point (PEP), while the actual components of the infrastructure rely on other implementations rather than XACML.

\paragraph{\textbf{Example.}}

The first scenario illustrates the need for the controllers of self-adaptive authorisation infrastructures to be aware of their limitations.
Once the compromised service has been detected, the self-adaptive authorisation infrastructure should synthesise a plan that would remove the
service's credentials from the authorisation policy, and would reload the new policy into the authorisation infrastructure.
If PERMIS is used as an authorisation server, for example, policy cannot be reloaded without restarting the entire server.
A plan that requires reloading a policy would fail, unless it allows for the server to be restarted as an alternative.

\subsubsection{Infrastructure Boundary}
\label{sec:plan:synthesis:infrastructure-boundary}

In a federated authorisation infrastructure, some components can be managed by third parties, and this should be captured by control boundaries that can be dynamic according to the role of the components of the infrastructure.
These components may have different or even conflicting goals, hence negotiations between components are needed in order maximise their self-interest.
In a self-adaptive authorisation infrastructure, the controller may only have partial control, or no control at all, over some of the components of the federated  authorisation infrastructure.
Considering that a controller needs to act on some components of the infrastructure that are owned by a third party, it is necessary that each party can trust each other in order to enable the negotiations.

\paragraph{\textbf{Challenge.}}

In a federated authorisation infrastructure in which self-adaptation underpins the authorisation services, the challenge is the ability of establishing boundaries of awareness and influence, and the ability of handling the dynamic nature of these boundaries.

\paragraph{\textbf{Relevance.}}

Control boundaries are particularly relevant in federated
authorisation infrastructures, where the controller does not have
direct control over the third party components.
The controller may be able to request components of a federated authorisation framework  to enact some changes, but these changes may only be accepted if they do not conflict with the  goals of those components.
In a federated authorisation infrastructure, boundaries can be related to what can be monitored and control, and to levels of trust, for example.

\paragraph{\textbf{Example.}}

The third scenario illustrates the issue of control boundary.
If the subject deleting data at an alarming rate has authenticated using a
third party service outside of the controller's control boundary, the
controller may devise a plan that first asks the third party service
to ban the user for some time.
However, the controller cannot
\emph{force} the third party service to enact this ban.
Thus, the controller may also generate an alternative, which would
ban the third party identity service entirely for some time.

\subsection{Execute}
\label{sec:exec}


The Execute stage is responsible for executing the adaptation plan generated during the Plan stage.
However, the execution of the plan  may not always be straightforward since it involves several distinct needs, including the following ones:
\begin{itemize}
\item  meet the objectives of the adaptation plan (including, the synthesis of effectors, deployment of probes/gauges), and provide assurances that effectors have indeed carried out their actions (i.e., feedback of success);
\item   effectors must trust the controller of the self-adaptive authorisation infrastructure, in terms of authentication,  authorisation, and non-repudiation;
\item   Execute stage is able to coordinate the effectors in issues, like, concurrency, rollbacks and commits, recovery from failed plans, and heterogeneity of effectors;
\item   adaptation plans incorporate redundancies for making its execution more resilient, and for supporting the provision of trust that resilience can be achieved;
\item   execution of the adaptation plan  is secure in order to avoid exploitation of vulnerabilities;
\item   adaptation plans incorporate abstract commands since it should be down to the effector to decide how to implement a given action of the plan;
\item  synthesise and/or deploy probes, gauges and effectors, or even update its own adaptation strategy in order to respond to new threats being detected;
\item  ability to reloading adapted authorisation policies, or restarting authorisation services or other related services;
\item  ability to communicating with third-party identity providers, amongst other services, for example.
\end{itemize}

Underpinning all these needs, there is the fundamental need to provide assurances that the execution of the plan is according to its specifications. 
The execution of a plan may involve checking post-conditions, and it should provide feedback of its progress. 
In the following, some of the above needs will be detailed in term of challenges.

\subsubsection{Deployment and Withdrawl of Probes, Gauges and Effectors}
\label{sec:exec:probes-deployment}

Although the deployment and withdrawal of probes, gauges and effectors might be outside the context of Execute stage of a self-adaptive authorisation infrastructure, these are an integral part of the adaptation plan and its execution.
Probes, gauges and effectors could either be taken from a pool, which is  populated at development-time, or synthesised at run-time for allowing the system to react to unforeseen changes (see Section~\ref{sec:exec:effector-synthesis}).
This dynamic deployment and withdrawal  is different from the active monitoring challenge discussed in Section~\ref{sec:eng:monitoring:active}.
In active monitoring, deployed gauges and probes have their own self-adaptive mechanism.
In contrast, we discuss in this section the deploying and withdrawing of probes, gauges and effectors as the result of the evolution of self-adaptive authorisation infrastructures.
As such, the deployment of new probes, gauges and effectors instead of being under the direct responsibility of a self-adaptive authorisation infrastructure, we could have a higher level entity responsible for controlling the evolution of the self-adaptive authorisation infrastructure.

\paragraph{\textbf{Challenge.}}

One of the challenges in self-adaptive authorisation infrastructures is the ability to deploy and withdraw probes, gauges and effectors
because of the wide range, and volatile nature, of threats that the system has to protect itself against.

\paragraph{\textbf{Relevance.}}

In a self-adaptive authorisation infrastructure changes affecting the infrastructure or its environment should be handled by the controller, and the ensuing adaptations may have an impact on how the controller observes and effects the infrastructure and/or its environment.
Consequently, this may affect the probes, gauges and effectors that are deployed.
The complexity of the deployment can range from entirely pre-defined probes, gauges and effectors that simply need to be activated, to the synthesis
of new ones (see Section~\ref{sec:exec:effector-synthesis}).
An intermediate solution would be the ability to configure pre-defined probes, gauges and effectors for a particular use.

\paragraph{\textbf{Example.}}

The first scenario provides an example of probe deployment and withdrawal.
When the couple subscribes to the energy saving service, it is necessary to monitor the service's behaviour through the
deployment of probes and gauges.
If the probes and gauges are simply pre-defined, they will already have been written for the particular
service chosen by the couple.  A more challenging option is for the
self-adaptive authorisation infrastructure to configure the probes and gauges, depending on what the service has access to, what the system wants to monitor, and the implementation details of the chosen service.

\subsubsection{Automated Synthesis of Probes, Gauges, and Effectors}
\label{sec:exec:effector-synthesis}


It is not possible for the developers, at development-time, to foresee all the possible threats that the system could face.
Too many of those depend on the environment in which the system operates, and this environment can change at any time.
The synthesis of new probes, gauges and effectors at run-time allows the system to react to changes in the system or its environment that would otherwise affect the resilience of the self-adaptive authorisation infrastructure.

\paragraph{\textbf{Challenge.}}


The challenge  of a self-adaptive authorisation infrastructure to react to changes that are not foreseen at development-time is to synthesise probes, gauges and effectors at run-time.

\paragraph{\textbf{Relevance.}}

Responding to threats that had not been foreseen during development-time is  essential for self-adaptive authorisation infrastructures.
Occasionally, handling some of these threats may require probes, gauges and effectors that are not yet available.
In particular, the self-adaptive authorisation infrastructure may require new probes or gauges for monitoring new features of the system or its environment.
The ability to synthesise probes, gauges and effectors  during run-time will broaden the range of unforeseen threats that the system can protect itself against.


\paragraph{\textbf{Example.}}

To demonstrate the automated synthesis of effectors, we postulate the
deployment of a new identity provider.
Such a deployment could be part of the response to the compromised service described in the first scenario.
The identity provider that was used to carry out the
initial attack might have been banned, and a new one may need to be
deployed.
The deployment of a new identity provider requires the
self-adaptive authorisation infrastructure to deploy new effectors to
communicate with the new identity provider, e.g., to ask it to ban a
particular user for a given amount of time.
It may also be necessary to synthesise new probes if the infrastructure needs to monitor the new identity provider.

\subsubsection{Trust}
\label{sec:exec:trust}

Trust is necessary between the parties in a federated authorisation infrastructure.
The controller should trust that the other parties will carry out the plan as expected, and the other parties must trust that the controller acts in their advantage.

\paragraph{\textbf{Challenge.}}


A a key challenge in a self-adaptive authorisation infrastructure is to maintain trust between the parties by ensuring all parties behave as agreed.
This is not restricted to techniques that ensure trust is maintained, but also associated with strategies that are used when reacting to a breach of trust.


\paragraph{\textbf{Relevance.}}

If a malicious user has been detected and reported to the identity service, but the identity service fails to take action to suspend the malicious user, then the trust between the authorisation infrastructure and the identity service should be reevaluated.
As authentication is a critical component in access control.
It is crucial for the authorisation infrastructure to be able to react to such breaches of trust, as they may harm the protected system.

\paragraph{\textbf{Example.}}
The second scenario provides a good example regarding trust.
In this scenario, the authorisation infrastructure detects that the energy company is acting against the users' interest by attempting to collect too much data, and too often.
The self-adaptive authorisation infrastructure is able to detect this as a breach of trust.
The controller reacts by requesting to the identity services to revoke the credentials of the energy company, which should affect its ability to gather any further data.


\subsubsection{Update or Redeployment of Policies, and Sessions}
\label{sec:exec:update-vs-redeployment}

Self-adaptive authorisation infrastructures can adapt authorisation policies in
various ways.
The adaptation could either take the form of an update
of the current policy, where the controller sends the modifications  to the policy decision point (PDP).
Alternatively, the controller may create a whole new policy, and instruct the PDP to deploy it instead of the previous one.

When adapting authorisation policies, the infrastructure should also
consider the sessions that are currently open by a particular user, and
which may carry permissions that the user should not be assigned
anymore sessions.
All sessions could be revoked every time adaptation occurs.
Alternatively, sessions could be amended in order to reflect
the changes made in the authorisation policy.

\paragraph{\textbf{Challenge.}}

During the execution of the plan, the challenge is to reduce the system vulnerability while policies are updated or redeployed.

\paragraph{\textbf{Relevance.}}

The adaptation of an authorisation policy is an important part of a
self-adaptive authorisation infrastructure's reaction to an internal
threat.
It is important that the adaptation is completed in a timely
manner, in order to minimise the amount of time during which an
attacker can cause damage to the system.
The choice between updating
the existing policy or deploying a new one should take into account
the amount of time required for the new policy to be effective.
Updating an existing policy requires the controller to
communicate to the PDP only the changes to be made to the current
policy.
It is then the PDP's responsibility to enact those changes.
Deploying a new policy, however, requires the controller to
prepare a complete, updated policy, and to communicate in to the PDP,
which only needs to deploy it to replace the previous one.

Existing sessions should also be taken care of in order to adapt the
permissions given to the users that are logged on to the system while
the adaptation takes place.
Terminating all sessions is a simple
solution, but it will require each use to authenticate again.
In some
circumstances this may not be ideal, especially if adaptation occurs
often.
The alternative is to modify user sessions at run-time, which may be more difficult to implement.

A similar challenge could be associated with the deployment of new probes and effectors, in particular those that are third party. 
The heterogeneity and the inflexibility of such devices may introduce vulnerabilities during adaptation. 

\paragraph{\textbf{Example.}}

The third scenario illustrates the importance of handling existing
sessions when updating or redeploying the authorisation policy.
In this scenario, data is being deleted by an attacker using stolen,
legitimate credentials.
Once the attack has been detected by the Analyse stage, and the policy adapted by the Execute stage, the
session used by the attacker must be taken care of.
Depending on the
infrastructure's implementation, two alternatives are available.
First, the session can be terminated entirely, forcing the attacker to
attempt to authenticate again using the stolen credentials.
Second,
the privileges afforded to the session may be reduced, in such a way
that the attacker would be unable to continue deleting data, but
without terminating the session.

\subsubsection{Redundancy}
\label{sec:exec:redundancy}

Introducing redundancy in the Execute stage is one way to increase
the resilience of the system.
In case an effector fails to properly
execute a portion of the plan, the plan should incorporate redundancies, using other effectors, alternative solutions, or workarounds, that would allow to tolerate the failed execution.

Whilst the incorporation of redundancies in case of failure may be, in
part, the responsibility of Plan stage, it is the responsibility of the
Execute stage to monitor the execution of that plan, and to trigger the redundancy measures when necessary.

\paragraph{\textbf{Challenge.}}


The challenge is to  include sufficient redundancies in the execution of the plan in order to make adaptation more resilient against potential threats, being these either internal attacks or faults.

\paragraph{\textbf{Relevance.}}

The consequences of a failure during the execution of a plan while
adapting an authorisation policy can be disastrous.
At best, the old policy will still be in place, and the system will not be protected
against the newly identified threat.
At worst, the authorisation
infrastructure could fail, either by locking all users out, or by
letting everyone access everything.
Implementing redundant mechanisms
to update policies will increase the service's resilience.

\paragraph{\textbf{Example.}}

All the scenarios in the case study could provide an example of the
importance of reacting to failures during Execute stage.
In the second scenario, the inability of the infrastructure to reduce the
energy company's rate of data collection will directly harm the users'
privacy.
The plan can include contingency provisions that handle this
particular issue, and an alternative strategy, which is an integral part of the plan being executed, can then be deployed,
such as, giving the energy company empty readings when they try to collect data too often.


\subsection{Models}
\label{sec:models}

There are several types of models relevant to authorisation
infrastructures, as well as, many ways of using them for
self-adaptation.
In this section, we first categorise models for authorisation infrastructures into four types: authorisation polices, access control, threat, and adaptation.
We then focus on the challenges that stem from the use of these models, which include: portability, facilitating negotiation, history of models, uncertainty and conflicts in models, and model drift.


The examples in this section differ from those in the previous
section.
Instead of using the scenarios from our case study to
highlight challenges, we will instead discuss the case study in
general.
This is because models are the same in the application, and
do not depend on specific scenarios.

\subsubsection{Modelling authorisation policies}
\label{sec:models:authz-policies}

Authorisation policies are a central component of authorisation
infrastructures, where rules or assignments are defined, and whose
evaluation determines whether requests for access to protected assets
are granted or denied.
Policies can be very large, and they can be
distributed across several documents, using various technologies, or,
in the case of federated authorisation infrastructures, under the control of different
entities.

\paragraph{\textbf{Challenge.}}

Models of authorisation policies should be understandable yet precise for facilitating their manipulation by both the controller and system administrators.


\paragraph{\textbf{Relevance.}}

For supporting the automated manipulation of models, these need to be precise, and at the same time, in order to allow the validation of the models against their respective policies, these models need to be accessible to users. 
The reason being that, since authorisation policies determine the rendering of access control decisions, the authorisation policies models should be consistent with the actual authorisation policies in order to avoid discrepancies between the authorisation infrastructure and its controller.
So whatever changes are made on the authorisation policies, either by users or automated tools, these need to be accurately reflected on their corresponding models. 


\paragraph{\textbf{Example.}}

In our case study, the Customer Energy Management  System (EMS) maintains the authorisation policies.
In order to present a coherent view of the policies to the user, the EMS must be able to collect all the policies, which may be
expressed in different languages or using different access control models, and collate them in a way that is easy for users to make sense of them.
Furthermore, since third parties or service providers, such as the energy provider, may have some control over parts of the authorisation policies, their correspondent models should be able to capture any changes made on the policies. 
On the top of this, there is the controller, which should be able to manipulate the different control models.


\subsubsection{Modelling access control}
\label{sec:models:system-arch}

Authorisation infrastructures often involve several key components that are connected for rendering access control decisions.
For example, the XACML standard recommends that an authorisation infrastructure should be separated into the following distinct components~\cite{xacml}: Policy Decision Points (PDP), Policy Enforcement Points (PEP), Policy Information Points (PIP), and Policy Administration Points (PAP).
Architectural models at the controller should be able to capture the components of an authorisation infrastructure, and how these components are connected.


\paragraph{\textbf{Challenge.}}

Architectural models should be dynamic because the infrastructure is expected to change, and these should be capture by the models. 
Such dynamic architectural models should be able to capture issues, such as, the unavailability of components.
In the context of federated authorisation infrastructures, architectural models should also be able to support access control and provide assurances of this ability. 

\paragraph{\textbf{Relevance.}}

In self-adaptive authorisation infrastructures, dynamic architectural models are essential for enabling the controller to handle changes affecting the infrastructure.
Architectural models enable to analyse the consequence of threats to the infrastructure, and investigate potential architectural solutions for mitigating those threats.
For example, if an identity service fails to revoke credentials from subjects that are perceived as persistent attackers, the self-adaptive authorisation infrastructure may chose to disconnect that identity service from its infrastructure.
However, before implementing that solution, the authorisation infrastructure may evaluate the impact of such a measure towards its users.

\paragraph{\textbf{Example.}}

In our case study, the EMS is at the centre of the federated authorisation infrastructure, and its architecture should include the third party service provider and the energy provider, both sharing the EMS control over the smart meter.
When adaptation occurs, the self-adaptive authorisation infrastructure must be aware of all those components in order to effect changes across the infrastructure.

\subsubsection{Modelling threats}
\label{sec:models:threats}

The purpose of self-adaptive authorisation infrastructures is to defend the system against threats.
Since threats can change throughout the life time of an infrastructure, threat models should be dynamic, i.e., models that are able to change according to the threats to the infrastructure.

\paragraph{\textbf{Challenge.}}

For reasoning about threats in self-adaptive infrastructures, a modelling language for expressing dynamic threat models is needed.
In addition to representing threats, threat models should capture the likelihood of their occurrence, the potential harm they can cause to the infrastructure's assets, and which countermeasures can be taken to address them.
If the self-adaptive infrastructure is capable of discovering previously unknown threats, then the threat model should be adaptable at run-time.

\paragraph{\textbf{Relevance.}}

Threats, whether internal or external, are what self-adaptive authorisation infrastructures try to defend the system against. 
This can only be done if these infrastructures have a suitable model of threats, and if these models cannot be adapted according to ever changing threats a vulnerability will ensue.

\paragraph{\textbf{Example.}}

A threat model for our case study could represent, for example,  abuse regarding smart meter's queries. 
A possible threat and its response is discussed in the second scenario. 
A model of the threat is necessary for the infrastructure to realise that an attack is taking place. 
The model could simply be a counter representing the number of queries
from the energy retailer over a set period of time, with a threshold
above which the attack is deemed to happen. 
A more complex model could
maintain a historical distribution of the queries' frequency, together
with a variation limit that, once reached, would represent an attack
taking place.

\subsubsection{Modelling adaptation}
\label{sec:models:adaptation}

The various models supporting self-adaptive authorisation infrastructures should take into account the run-time adaptation capabilities of the infrastructure.
In particular, adaption models should represent what parts of the authorisation infrastructure can be adapted, how the adaptation can be executed and in which order, and when or under which conditions adaptation can happen.
\paragraph{\textbf{Challenge.}}

There is the need to specify adaptation models that would be able to coordinate adaptations taking at different levels of an authorisation infrastructure, and to identify what kind of assurances those models can provide.

\paragraph{\textbf{Relevance.}}

Considering that both authorisation infrastructures and their environments are intrinsically dynamic, adaptation models should be related to the architectural models of the infrastructure, models of the authorisation policies, and threat models. 
Adaption models should also capture how the controller communicates with the authorisation infrastructure, which includes the monitoring and controlling of the infrastructure. 
If adaptation models do not related to all models that enable the support of self-adaptation of authorisation infrastructures then inconsistencies might arise regarding the how self-adaptation is enacted.

\paragraph{\textbf{Example.}}

In our example, adaptation models should be able to capture the different aspects of adapting an authorisation infrastructure, which includes adapting its architectural configuration or the components that are part of that configuration.
An example of component adaptation would be its ability of adapting authorisation policies in order to modify users' ability to perform certain operations, for example, restricting the amount of smart meter readings.
An adaptation model, that would be part of Customer Energy Management  System (EMS), should specify which parts of the authorisation policy can be adapted and how. 
It should also specify how the adapted policy can be deployed in the infrastructure.
For example, it might be the case that the component responsible for the authorisation service must be restarted, as it is not possible to reload the authorisation policy without restarting the component.

\subsubsection{Portability}

Portability can take different forms in self-adaptive authorisation infrastructures.
It could mean that the system can be deployed on various types of infrastructures.
It could also mean that the system should be able to function in an heterogeneous environments, where
different subsystems can communicate with each other.
The former requires some form of vertical transformation, where abstract elements can be concretised in various ways, depending on the underlying
infrastructure. 
The latter requires some form of horizontal transformation, where models and data can be communicated between
subsystems that use different representations.

\paragraph{\textbf{Challenge.}}

The need for developing self-adaptive authorisation infrastructures that can be vertically and horizontally portable.
Vertically portable self-adaptive authorisation infrastructures can easily be redeployed on different environments or implementations of components. 
Horizontally portable self-adaptive authorisation infrastructures have components that may use different data formats and protocols, but are still able to communicate with each other.

\paragraph{\textbf{Relevance.}}

Federated authorisation infrastructures have components controlled by various entities. 
system they use. 
Components of these infrastructures may run different technologies, or different versions of the same technology. 
They may also evolve and change at any time. 
Therefore, it is necessary for the self-adaptive authorisation infrastructure to be designed in a portable way.

\paragraph{\textbf{Example.}}

In our case study, the self-adaptive authorisation infrastructure may be federated, with different entities handling parts of the authorisation and/or authentication infrastructure. 
For example, the owner may delegate the authentication to OAuth providers, such as,  Google or Twitter, while the household owners and the energy provider may have share responsibility over authorisation for the access to the smart meter readings.

\subsubsection{Facilitating negotiation}

Federated authorisation infrastructures require several components working together. 
However, these components may be owned and controlled by different entities, who may have conflicting goals and interests. 
Therefore, it may be necessary for multiple components to negotiate a solution that satisfies all parties.

\paragraph{\textbf{Challenge.}}

Models are required to facilitate negotiation between several components of a federated authorisation infrastructure. 
In self-adaptive authorisation architectures, these models should allow components to understand the consequences of the proposed changes on users' ability to use the system, and should allow components to express agreement or disagreement with some of those changes.

\paragraph{\textbf{Relevance.}}

Self-adaptive federated authorisation infrastructures must be able to handle
negotiation between several of their components. 
They must be able to
exchange proposed solutions to problems, and indicate agreement or
preferences. 
The solutions will contain models of the proposed
changes, in order for the components to make informed decisions about
the proposals they are presented with.

\paragraph{\textbf{Example.}}

In our case study, the EMS and the energy company share access to the
smart meter readings. 
It is in the household's interest to minimise
the energy company's data collection frequency, and the amount of data
it can collect. 
However, there is a minimum amount of data that the
energy company needs to collect in order to guarantee its service, and
to be able to correctly invoice the household. 
Since both the
household's and the energy company's requirements can change, they may
have to re-negotiate the data collection frequency. 
A suitable model
for such a negotiation would represent the amount of data that can be
collected and its frequency, and would allow each party to express
agreement or disagreement, in whole or in part, and to propose
alternatives until a solution can be found.

\subsubsection{Capturing the history of models}
\label{sec:models:history}

Capturing the history of models allows for the analysis of changes that happened in the past. 
The detection of long-running attacks, forensics analysis, and the detection of the entry point of an
attacker all require access to historical data about the state of the
system.

\paragraph{\textbf{Challenge.}}

Self-adaptive authorisation infrastructures should be able to keep a
history of the models that they maintain, in such a way that does not
degrade performance, yet allows for efficient analysis of past
events. 
The ability to correlate changes to different models is
especially important.

\paragraph{\textbf{Relevance.}}

Attacks can be carried out over long periods of time. 
Hence,
understanding them may require to analyse the past states of the self-adaptive authorisation
infrastructure, both in order to detect and prevent attacks, but also
in order to identify patterns of suspicious behaviour over a long
period of time.

\paragraph{\textbf{Example.}}

The infrastructure in our case study focuses on the detection of
internal threats, which means that, before carrying out their attacks,
the attacker must have been authenticated. 
In order to find out whether an attack was enacted by the legitimate holder of the credentials in question, or by an external attacker who managed to impersonate them,
it requires analysing the state of the infrastructure at the time of authentication, how the authentication was carried out, as well as, compare the behavioural patterns of the user at different points in time.

%
%
%
%
%
%
%

\subsubsection{Analysis capabilities}

The Analyse Stage of a self-adaptive system depend on the available models in the knowledge base. 
If these models cannot completely reflect the reality they represent, the Analyse stage may not be able to always come to the correct conclusion, or even come to a conclusion at all. 
Therefore, analysis on incomplete models may lead to uncertainty. 
If several analysis components are used, this may also lead to conflicting results.

\paragraph{\textbf{Challenge.}}

Self-adaptive authorisation infrastructures should be able to deal with uncertainty and conflicts, and these may need to be encoded in the models.

\paragraph{\textbf{Relevance.}}

Detection of insider threats is difficult because a smart malevolent insider will attempt to try and pass their usage as legitimate. 
Therefore, there is often no clear-cut distinction between legitimate and malicious users, making their detection difficult and ambiguous. 
Moreover, in a federated self-adaptive authorisation infrastructure, various Analyse stages may reach different conclusions, even when considering the same data.

\paragraph{\textbf{Example.}}

The authorisation infrastructure may use models of users' access to resources to detect insider threats. 
One such model may represent each access request, its
timestamp, the resources requested, and the authorisation infrastructure's response. 
One analysis  may use the number of failed requests
per hour, and classify users as legitimate in the number of failed
requests is smaller than a threshold $x$, suspicious if between $x$
and $y$ (with $x < y$), and a threat if the number of failed requests
is higher than $y$. In this case, suspicious users are an uncertain
result: the analysis  was not able to give a definite answer.

Another analysis  could use the same data, but perform
different operations to detect insider threats. 
It could look a the
\emph{total} number of requests, whether they were denied or not by
the authorisation infrastructure. 
Similarly to the first analysis component,
if a user produces less than $a$ requests per hour, it is considered
legitimate; between $a$ and $b$ ($a < b$) requests per hour, it is
considered suspicious, and over $b$ requests per hour, it is
considered a threat. 
It is possible that the two analysis 
will disagree on the nature of a particular user. 
The self-adaptive
authorisation infrastructure needs to deal with this conflict. 
For
example, it could run more analysis components over the same data
until it finds sufficient confirmation, or it could take a
conservative approach and treat the user as a threat if at least one
the analyses has identified it as a threat.

\subsubsection{Model drift}
\label{sec:models:drift}

Dynamic models are at risk of drift over time.
Model drift is the
progressive increase in the discrepancy between the model and what the
model represents.
In self-adaptive authorisation infrastructures, dynamic models are used to
represent the target system, as well as its environment.
If the models
do not correctly reflect the reality, this may lead to sub-optimal, or
harmful, adaptation decisions.

\paragraph{\textbf{Challenge.}}

In self-adaptive authorisation infrastructures, model drift should be avoided in order to
reliably detect suspicious activity and identify malicious actors.


\paragraph{\textbf{Relevance.}}

Attackers are likely to try and masquerade their actions as legitimate
in order to escape security measures.
Therefore, it is important for
self-adaptive authorisation infrastructures to keep models that are
very close to reality - even a small drift may be used by the attacker
to cover their tracks.

\paragraph{\textbf{Example.}}

In our case study, the self-adaptive authorisation infrastructure maintains a model of the
target system, which may include distributed access control policies
over several components of the  authorisation infrastructure.
If the self-adaptive authorisation infrastructure cannot keep an accurate representation of these policies, then it may
believe that an authorisation request that the target system will
accept would be rejected, or vice-versa, leading to incorrect
adaptation decisions.
For example, the self-adaptive authorisation infrastructure may recognise
that a user who is acting suspiciously has less permissions than he
actually has, and decide to treat it as a low priority threat, where
in reality, the user's  level of access should warrant a hight priority
resolution of the issue. 




\section{Conclusions}
\label{sec:conclusions}

\marginpar{\tiny RDL - evaluate the appropriatness of the MAPE-K loop for supporting self-adaptive access control.}

The provision of self-adaptive authorisation infrastructures is a promising solution to protect systems against the dynamic nature of attacks and uncertainties associated with them, such as, insider threats. 
In this paper, we have presented how this could be achieved architecturally by separating the specification of  policies (i.e., self-adaptive authorisation) from the enforcement of these policies (i.e., self-adaptive access control).
We have also presented several technical challenges associated with the self-adaptation of authorisation infrastructures, which followed the stages of the MAPE-K feedback control loop. 
Each of the technical challenges was presented in terms of their relevance, and an example was provided for demonstrating their pertinence.
Of course, the list of technical challenges was not exhaustive since several of them were not included in the description due to space constraints. 
Moreover, in addition to the identified technical challenges, restricted by our experience in building and deploying self-adaptive of authorisation infrastructures~\cite{Bailey:2014,baileyPhD2015},  one would expect new technical challenges to arise, depending on authorisation infrastructure and their deployment. 

For presentation purposes, it was natural to follow the  MAPE-K feedback control loop for identifying the technical challenges, however, questions may be asked about the appropriateness of MAPE-K loop when finding solutions to the wide range of identified challenges. 
Authorisation infrastructure are inherently quite complex infrastructures, which can be geographically distributed, and this might require architectural solutions  for the controller that might go beyond what the classical MAPE-K loop can offer~\cite{deLemos2017,Weyns:2012:FUR:2168260.2168268}.
For example, if perpetual evaluations~\cite{Weyns2017} are needed in order to obtained assurances of the confidentiality, integrity, and availability of computer based resources regarding the adaptations performed to the authorisation infrastructure, then new ways of enforcing separation of concerns are needed at the controller level.
This of course will raise a new set of technical challenges that should be specific to the provision of assurances. 

\bibliographystyle{abbrv}
\bibliography{ref}

\end{document}